\documentclass[aps, prb, twocolumn, amsmath, amssymb, showpacs, superscriptaddress]{revtex4-1}

\usepackage[utf8]{inputenc}
\usepackage{graphicx}
\usepackage{units}
\usepackage{color}
\usepackage[pdftex, colorlinks=true, linkcolor=myblue, citecolor=myblue]{hyperref}
\usepackage[varg]{txfonts}

\definecolor{myblue}{rgb}{0.1,0.,.85}

\begin{document}

\title{Euler buckling instability and enhanced current blockade
in suspended single-electron transistors}

\author{Guillaume Weick}
\affiliation{Institut de Physique et Chimie des Mat\'eriaux de Strasbourg (UMR
7504), CNRS and Universit\'e de Strasbourg, 23 rue du Loess, BP 43, F-67034
Strasbourg Cedex 2, France}

\author{Felix von Oppen}
\affiliation{Dahlem Center for Complex Quantum Systems \& Fachbereich Physik, 
Freie Universit\"at Berlin, 
Arnimallee 14, D-14195 Berlin, Germany}

\author{Fabio Pistolesi}
\affiliation{Centre de Physique Mol\'eculaire Optique et Hertzienne (UMR 5798), 
CNRS and Universit\'e de Bordeaux I, 351 Cours de la Lib\'eration, F-33405
Talence Cedex, France}
\affiliation{Laboratoire de Physique et Mod\'elisation des Milieux Condens\'es
(UMR 5493), CNRS and Universit\'e Joseph Fourier, 25 avenue des Martyrs, BP 166,
F-38042 Grenoble Cedex, France}

\begin{abstract}
Single-electron transistors embedded in a suspended nanobeam or carbon nanotube
may exhibit effects originating from the coupling of the electronic
degrees of freedom to the mechanical oscillations of the suspended structure.
Here, we investigate theoretically the consequences of a capacitive
electromechanical interaction when the supporting beam is brought close to the
Euler buckling instability by a lateral compressive strain.
Our central result is that the low-bias
current blockade, originating from the electromechanical coupling for the 
classical resonator, is strongly enhanced near the Euler instability. We predict that the
bias voltage below which transport is blocked increases by orders of magnitude for typical
parameters. This mechanism may make the otherwise elusive classical current
blockade experimentally observable.
\end{abstract}


\maketitle

\section{Introduction}
\begin{figure}[tb]
\includegraphics[width=.82\columnwidth]{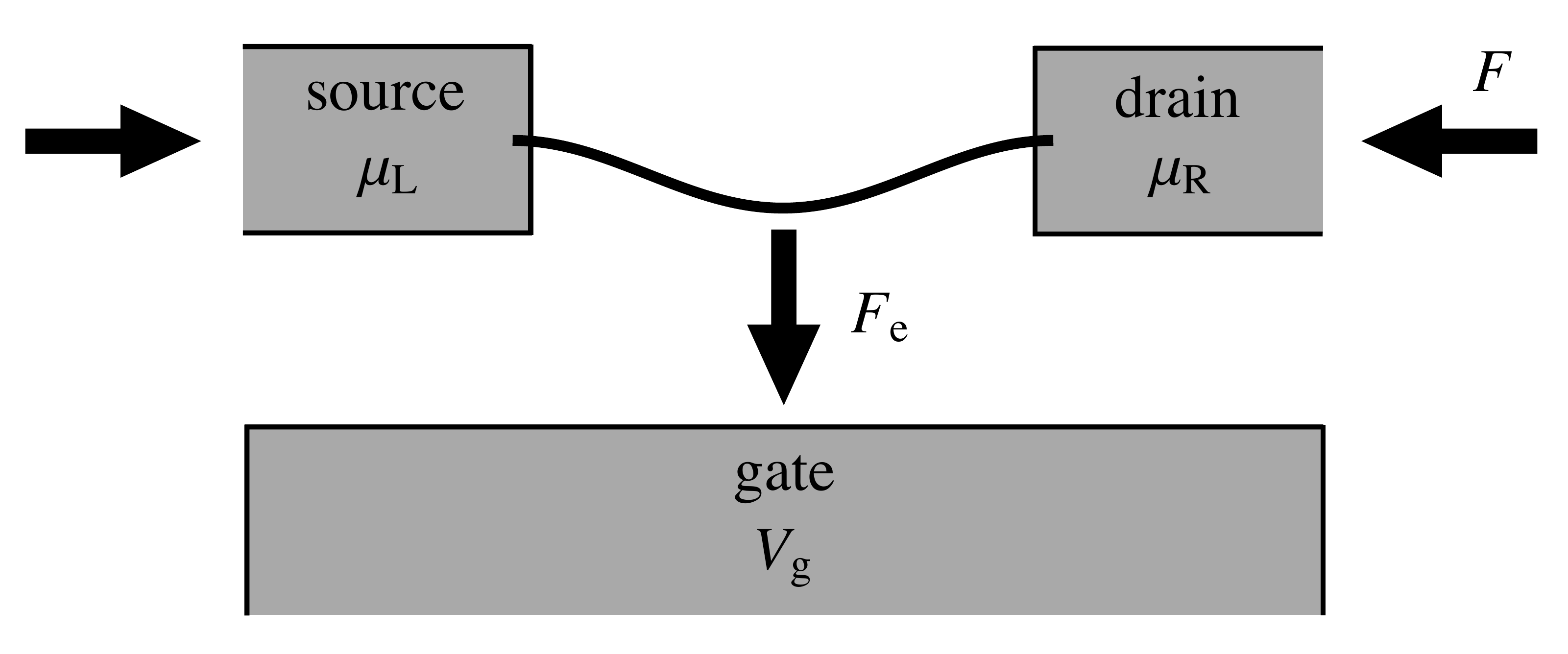}
\caption{\label{fig:SET}%
Sketch of the considered system: a suspended doubly-clamped beam
forming a quantum dot electrically connected to source and drain 
electrodes held at chemical potentials $\mu_\mathrm{L}$ and $\mu_\mathrm{R}$ by
the bias voltage $V$, respectively. The beam is capacitively coupled to a metallic 
gate kept at a voltage
$V_\mathrm{g}$, which induces a force $F_\mathrm{e}$ that attracts the beam
towards the gate electrode.
An additional, externally controlled compressional force $F$ acts on the beam and induces a
buckling instability.}
\end{figure}

Single-electron transistors (SET) are extremely sensitive devices, and 
are investigated as position detectors for 
nano-electromechanical systems (NEMS).\cite{knobe02_APL, knobe03_Nature,
armou02_PRL} 
But the reduced size of the mechanical resonator implies that the back-action of the 
SET can have significant effects on the mechanical degree of freedom, 
such as the generation of self-oscillations.\cite{gorel98_PRL, pisto05_PRL,
usman07_PRB}
In practice, the detector and the resonator have to be investigated collectively as a single 
system. 
A prominent example of the new effects displayed by this device is the current blockade 
that appears at low bias voltage $V$ when the SET is coupled capacitively to a
classical oscillator. \cite{pisto07_PRB} 
The physical idea behind this phenomenon is simple: The presence of an extra electron on the central island of the 
SET induces an additional
electrostatic force $F_\mathrm{e}$ on the oscillator (see Fig.~\ref{fig:SET}). 
The equilibrium position of the oscillator is thus shifted by a distance $F_\mathrm{e}/k$, where $k$ is the oscillator 
spring constant. 
After such a displacement, the gate voltage $V_\mathrm{g}$ seen by the SET
changes by a quantity 
of the order of  $F_\mathrm{e}\times F_\mathrm{e}/ek \equiv E_\mathrm{E}/e$, where $e$ is the electron charge. 
The dimension of the conducting window in $V_\mathrm{g}$ is controlled by $V$, 
since at low temperatures
current can flow through the device only if $|V_\mathrm{g}|\lesssim V$
(when measuring $V_\mathrm{g}$ from the degeneracy point).
Thus, for $eV<E_\mathrm{E}$, the fluctuation of the
electronic occupation of the central island suffices to bring the device out of the conducting 
window. 
The current is blocked for $eV<E_\mathrm{E}$ and a mechanical bistability
appears. \cite{pisto07_PRB} 
This phenomenon is the classical counterpart of the Franck-Condon blockade in
molecular devices \cite{koch05_PRL, koch06_PRB}
that has recently been observed in suspended carbon nanotubes for high-energy 
vibrational modes.\cite{letur09_NaturePhysics}
The classical case has been theoretically studied in the case of a single-level
quantum dot, \cite{mozyr06_PRB, pisto08_PRB}
as well as in the metallic case.\cite{armou04_PRB, blant04_PRL, doiro06_PRB,
pisto07_PRB}

Recent experiments \cite{steel09_Science, lassa09_Science} on suspended carbon nanotubes have observed a reduction
of the mechanical resonance frequency of the fundamental bending mode at low
bias voltages
and for $V_\mathrm{g}$ near the degenerate region. 
This effect is a precursor of the mechanical instability and thus of the current blockade.
But the complete observation of the latter phenomenon is difficult since the typical
value of $E_\mathrm{E}$ is only of a 
few $\mu$eV, thus smaller than cryogenic temperatures. 
In order to increase $E_\mathrm{E}$, one can increase the electrostatic coupling between the 
oscillator and the SET since $E_\mathrm{E}$ depends quadratically on
$F_\mathrm{e}$.
But another way of strengthening the effect would be to reduce the spring constant
$k$ of the oscillator. 
The reason is that softer oscillators will displace more under the influence of the 
electrostatic force $F_\mathrm{e}$, and thus will see a larger change in 
the gate voltage when electrons tunnel in.
A way of reducing $k$ in a controlled manner is to operate a doubly-clamped
beam subject to a lateral compression force $F$. The latter can bring
the beam to the well-known
Euler buckling instability \cite{euler, landau} (see Fig.~\ref{fig:SET}). 
Under the action of the force $F$, the system exhibits a continuous transition from a flat 
to a buckled state, while the fundamental bending mode becomes softer as one
approaches the mechanical instability 
($k\rightarrow 0$).
It is clear that this does not imply a divergence of $E_\mathrm{E}$, since 
at the transition anharmonic terms will modify 
the simple arguments given above. However, a strong enhancement of 
$E_\mathrm{E}$ is expected.
The Euler instability has been studied both experimentally \cite{falvo97_Nature,
carr03_APL, carr05_EPL, roode09_APL} and
theoretically \cite{carr01_PRB, werne04_EPL, peano06_NJP, savel06_NJP} in
micro- and nanomechanical systems.
We have recently considered the Euler instability in NEMS for the case where 
$F_\mathrm{e}$ is negligible with respect to the intrinsic electron-phonon
coupling. \cite{weick10_PRB} 

In this paper we investigate in detail the idea of increasing the current
blockade by exploiting the Euler instability, considering how the anharmonic
terms, the temperature, and the 
non-equilibrium fluctuations modify the simplified picture given above.
We find that near the buckling instability, the current blockade induced by the
mechanical resonator is strongly enhanced, rendering this effect experimentally
observable.

The paper is structured as follows: 
In Sec.~\ref{Model} the model 
used to describe the system is introduced. 
In Sec.~\ref{sect:Fokker}, a statistical description
in terms of a Fokker-Planck equation is given. It is then  
used in the remainder of the paper to determine the current and 
the mechanical behavior of the system.
In Sec.~\ref{sect:mean-field}, the enhancement of $E_\mathrm{E}$ is obtained 
at mean-field level.
We discuss the effects of thermal and charge fluctuations 
on the results in Sec.~\ref{sect:temperature}. In Sec.~\ref{sec:average_charge},
we investigate the consequences of a finite average excess charge on the quantum
dot for our results.
Sec.~\ref{sect:experiment} presents some estimates of the 
effect we are predicting for recently-realized experiments.
We conclude in Sec.~\ref{sect:conclusions}. We relegate 
several technical issues to appendices.

\section{Model}
\label{Model}
As a representative model for the problem outlined in the
Introduction, we 
consider a quantum dot embedded in a doubly-clamped beam 
as shown in Fig.~\ref{fig:SET}. 
The presence of the metallic gate near the dot is responsible for the 
coupling of the bending modes of the beam to the 
charge state of the dot.
The Hamiltonian of the system can then be written as 
\begin{equation}
H=H_\mathrm{vib}+H_\mathrm{SET}+H_\mathrm{c}, 
\end{equation}
where $H_\mathrm{vib}$ describes the oscillating modes of the nanobeam, 
$H_\mathrm{SET}$ the electronic degrees of freedom of the 
single-electron transistor, and $H_\mathrm{c}$ the coupling between the
SET and the resonator.
The model describes, for instance, transport through suspended carbon nanotubes 
as considered in the experiments of Refs.~\onlinecite{letur09_NaturePhysics},
\onlinecite{steel09_Science}, \onlinecite{lassa09_Science}, and 
\onlinecite{hutte09_NL}.
Notice that the model also describes an
alternative setup that may be experimentally realized, namely a non-suspended 
quantum dot coupled to a beam-like gate electrode to which a compressive strain is
applied.

Using standard methods of elasticity theory one can show that, 
close to the buckling instability, the frequency $\omega$ of the fundamental bending mode
of the nanobeam vanishes while those of the 
higher modes remain finite.\cite{landau} 
This allows one to retain only the fundamental mode parametrized by 
the displacement $X$ of the center of the beam.
As detailed in Appendix \ref{sec:Euler}, the Hamiltonian representing the oscillations of the nanobeam thus takes the Landau-Ginzburg 
form \cite{carr01_PRB, carr03_APL, werne04_EPL, peano06_NJP, weick10_PRB}
\begin{equation}
\label{eq:H_vib}
     H_\mathrm{vib}=\frac{P^2}{2m}+\frac{m\omega^2}{2}X^2+\frac{\alpha}{4}X^4, 
\end{equation}
where $P$ is the momentum conjugate to $X$. 
For a doubly-clamped uniform nanobeam of length
$L$, linear mass density $\sigma$, and bending rigidity $\kappa$, 
one can show \cite{werne04_EPL, peano06_NJP} that close to the instability the effective mass of the beam is 
$m=3\sigma L/8$.
The fundamental bending mode frequency reads
\begin{equation}
\label{eq:omega}
\omega=\omega_0\sqrt{1-\frac{F}{F_\mathrm{c}}},
\end{equation}
where $F$ is the compression force, 
$F_\mathrm{c}=\kappa(2\pi/L)^2$ the critical force at which buckling
occurs, and $\omega_0=\sqrt{\kappa/\sigma}(2\pi/L)^2$. 
The positive parameter $\alpha=F_\mathrm{c}L(\pi/2L)^4$ ensures the stability of the system for
$F>F_\mathrm{c}$.\cite{footnote:CNT}
For $F<F_\mathrm{c}$ ($\omega^2>0$), $X=0$ is the only stable
solution, and the beam remains straight. 
For $F>F_\mathrm{c}$, it buckles into one
of the two metastable states at $X=\pm\sqrt{-m\omega^2/\alpha}$.
Notice that in
writing Eq.~\eqref{eq:H_vib}, we assumed that the nanobeam cannot rotate around its
axis due to clamping at its two ends.

Electronic transport is accounted for by
the SET Hamiltonian consisting of three parts,
\begin{equation}
H_\mathrm{SET}=H_\mathrm{dot}+H_\mathrm{leads}+H_\mathrm{tun},
\end{equation}
where 
$H_\mathrm{dot}$ describes the quantum dot, $H_\mathrm{leads}$ the left
(L) and right (R) leads, and $H_\mathrm{tun}$ the tunneling between leads and
dot. 
Explicitly,
\begin{equation}
H_\mathrm{dot}=\left(\epsilon_\mathrm{d}-e\bar V_\mathrm{g}\right)n_\mathrm{d}+
\frac{U}{2}n_\mathrm{d}(n_\mathrm{d}-1),
\end{equation}
with $n_\mathrm{d}=d^\dagger d$, and $d^\dagger$ ($d$) creates (annihilates)
an electron on the dot, 
$\bar V_\mathrm{g}=C_\mathrm{g}V_\mathrm{g}/C_\Sigma$, with $C_\mathrm{g}$ and $C_\Sigma$ the gate
and total capacitances of the SET, respectively. The intra-dot Coulomb repulsion
is denoted by $U$.
In the following, we set $\epsilon_\mathrm{d}=0$, measuring $V_\mathrm{g}$ from
the degeneracy point.
The left and right leads are assumed to be Fermi
liquids at temperature $T$ with chemical potentials $\mu_\mathrm{L}$ and
$\mu_\mathrm{R}$ (measured from $\epsilon_\mathrm{d}$), respectively. 
A (symmetric) bias voltage $V$ is applied to the
junction such that $\mu_\mathrm{L}=-\mu_\mathrm{R}=eV/2$. 
The lead Hamiltonian reads 
\begin{equation}
H_\mathrm{leads}=\sum_{ka}(\epsilon_k-\mu_a)c_{ka}^\dagger c^{\phantom{\dagger}}_{ka},
\end{equation}
with $c_{ka}$ the annihilation operator for a spinless electron of momentum $k$
in lead $a=\mathrm{L,R}$.
\cite{footnote:spinless}
Finally, tunneling is accounted for by the Hamiltonian 
\begin{equation}
H_\mathrm{tun}=\sum_{ka} (t_ac_{ka}^\dagger d+\mathrm{h.c.}),
\end{equation}
with $t_a$ the tunneling amplitude between the quantum dot and lead $a$.

Two different kinds of couplings exist between the electronic occupation of the dot $n_\mathrm{d}$ 
and the vibrational degrees of freedom:
(i) an intrinsic one that originates from the variation of the electronic
energy due to the elastic 
deformation of the beam, \cite{maria09_PRB} and 
(ii) an electrostatic one, induced by the capacitive coupling to the gate electrode of the SET. \cite{sapma03_PRB,
izumi05_NJP, sapma06_PRL, flens06_NJP}
By symmetry, the former is quadratic in the amplitude $X$ and its effect on the Euler instability has been considered in Ref.~\onlinecite{weick10_PRB}. 
The latter is linear in $X$ and here we are interested in the case where the
second coupling dominates over the first one.
Their relative intensity is controlled by the distance $h$ between the gate
electrode and the beam,  
since the intrinsic coupling does not depend on $h$, while the electrostatic force depends 
logarithmically on $h$.\cite{sapma03_PRB} Assuming that the beam is sufficiently
close to the gate electrode such that the capacitive coupling dominates, \cite{footnote:coupling}
we can write 
\begin{equation}
\label{eq:Hc}
  H_\mathrm{c}=F_\mathrm{e} X n_\mathrm{d}, 
\end{equation}
where $-F_\mathrm{e}$ is the force exerted on the tube when one excess electron
occupies the quantum dot (see Fig.~\ref{fig:SET}).
The model assumes that the gate voltage is such that only charge states with
$n_\mathrm{d}=0$ and 1 are accessible. For larger gate voltages overcoming the
charging energy of the quantum dot, the charge on the dot will instead
fluctuates between $N$ and $N+1$. This induces an additional constant force
bending the tube further. 
The effect of such a force on the classical current blockade will be discussed in
Sec.~\ref{sec:average_charge}.
Notice that in the case the suspended structure is the gate capacitance 
coupled to a static quantum dot, the intrinsic coupling is not present, and only
the capacitive electromechanical coupling has to be taken into account.

\section{Fokker-Planck description}
\label{sect:Fokker}
We are interested in describing the vicinity of the instability
where the relevant resonator frequency vanishes [see Eq.~\eqref{eq:omega}]. 
The mechanical degree of freedom can then be treated 
classically since for any reasonable temperature,
$\hbar \omega \ll k_\mathrm{B}T$.
The softening of the mechanical mode implies also a 
natural separation of timescales between the slow mechanical
mode and the fast electronic degrees of freedom, controlled 
by the typical tunneling rate $\Gamma$.
As detailed in Appendix \ref{sec:FP}, it is thus convenient to eliminate the fast modes and 
obtain a Fokker-Planck equation for the probability 
distribution $\mathcal{P}(X, P, t)$ of the slow 
mode:\cite{blant04_PRL,mozyr06_PRB, pisto07_PRB, weick10_PRB, husse10_preprint} 
\begin{align}
\label{eq:FP}
\partial_t\mathcal{P}=&
-\frac{P}{m}\partial_X\mathcal{P}-F_\mathrm{eff}(X)\partial_P\mathcal{P} 
+\frac{\eta(X)+\eta_\mathrm{e}}{m}\partial_P(P\mathcal{P})
\nonumber\\
&+\left(\frac{D(X)}{2}+\eta_\mathrm{e}k_\mathrm{B}T\right)\partial_P^2\mathcal{P}.
\end{align}
The effective force 
$F_\mathrm{eff}(X)=-\partial_XH_\mathrm{vib}+F_\textrm{c-i}(X)$
acting on the mechanical degree of freedom consists of two
parts: a force arising from the Hamiltonian \eqref{eq:H_vib} of the nanobeam,
$-\partial_XH_\mathrm{vib}=-m\omega^2 X-\alpha X^3$, and
a current-induced conservative force $F_\textrm{c-i}(X)=-F_\mathrm{e} n_0(X)$, 
proportional to the occupation of the dot averaged over a time long with respect to
$\Gamma^{-1}$, but short with respect to the period of the mechanical motion,
$n_0(X) = \langle n_\mathrm{d}\rangle_X$.
In Eq.~\eqref{eq:FP}, the diffusion constant $D(X)$
accounts for the fluctuations of the force associated with the coupling
Hamiltonian \eqref{eq:Hc} originating from the 
stochastic nature of the charge-transfer processes. 
Finally, retardation effects cause dissipation of the
mechanical energy, with damping coefficient $\eta(X)$.

To account for the quality factor $Q=m\omega_0/\eta_\mathrm{e}$ 
of the nanobeam mode, 
the mechanical degree of freedom 
is coupled to an additional environment at equilibrium (such as, e.g., a generic phonon
bath within the Caldeira-Leggett model \cite{weiss}), implying dissipation
and fluctuations controlled by an extrinsic damping
constant $\eta_\mathrm{e}$ entering Eq.~\eqref{eq:FP}. 
This extrinsic damping 
comes from several mechanisms coupling the mechanical mode to other degrees of
freedom: localized defects at the surface of the sample (thought to be the main
source of dissipation in semiconductor resonators \cite{mohan02_PRB} and which can be modeled
as two-level systems \cite{seoan07_EPL, seoan08_PRB}), clamping losses,
thermo-elastic losses, ohmic losses due to the gate electrode (which
have been predicted to be the dominant source of extrinsic dissipation for
graphene-based resonators in Ref.~\onlinecite{seoan07_PRB}), etc. Due to the
wide variety of these possible sources of extrinsic dissipation, we here assume for
simplicity that they can all be lump into the generic (Ohmic, memory-free
\cite{weiss}) damping constant $\eta_\mathrm{e}$.
Notice also that the phonon temperature of the bath
is typically lower, but of the same order as the
electronic temperature $T$. \cite{steel09_Science}  
For simplicity, we assumed in writing Eq.~\eqref{eq:FP} that both temperatures
coincide, as we do not expect a qualitative change of our results due to this
assumption.

The explicit form of the coefficients entering into the
Fokker-Planck equation \eqref{eq:FP} depends on the 
transport regime one considers (sequential
or resonant transport), as well as on the nature of the 
quantum dot (metallic or single-level quantum dot). 
In this paper we consider the case of a single level in the 
sequential tunneling regime, but 
a similar analysis can be carried out for the metallic (e.g., along the lines of
Refs.~\onlinecite{pisto07_PRB}, \onlinecite{blant04_PRL},
\onlinecite{weick10_PRB})
and the resonant transport regime (cf.\ Refs.~\onlinecite{mozyr06_PRB}, \onlinecite{pisto08_PRB},
\onlinecite{husse10_preprint}).
To be specific, we assume that $\hbar\Gamma=\sum_{a=\mathrm{L,R}}\hbar\Gamma_a\ll
k_\mathrm{B}T$,
with $\Gamma_a=2\pi|t_a|^2\nu/\hbar$ and $\nu$ the density of states at the Fermi
level of the leads.
We also assume the intra-dot Coulomb repulsion $U\rightarrow\infty$
such that double occupancy of the dot is forbidden. 
In this transport regime, the position-dependent rates for tunneling
into and out of the dot read \cite{elste08_APA}
\begin{align}
\label{eq:Gamma01}
\Gamma_{01}(X)&=\sum_{a=\mathrm{L,R}}\Gamma_a
f_\mathrm{F}\left(\frac{F_\mathrm{e} X-e\bar V_\mathrm{g}-\mu_a}{k_\mathrm{B}T}\right),
\\
\Gamma_{10}(X)&=\sum_{a=\mathrm{L,R}}\Gamma_a
\left[1-f_\mathrm{F}\left(\frac{F_\mathrm{e} X-e\bar V_\mathrm{g}-\mu_a}{k_\mathrm{B}T}\right)\right],
  \end{align}
respectively, where $f_\mathrm{F}(z)=(\mathrm{e}^z+1)^{-1}$ is the
Fermi function. 
Thus, the average occupation of the dot for a given mode amplitude $X$ is
\begin{equation}
\label{n0Gen}
n_0(X)=\frac{\Gamma_{01}(X)}{\Gamma},
\end{equation}
and, as shown in Appendix~\ref{sec:FP}, we have 
$D(X)=2F_\mathrm{e}^2n_0(X)[1-n_0(X)]/\Gamma$ and 
$\eta(X)=-F_\mathrm{e}\partial_Xn_0(X)/\Gamma$ for the current-induced diffusion and damping terms in
Eq.~\eqref{eq:FP}. \cite{blant04_PRL, footnote:fluctuation-dissipation}
The average current $I$ through the device can be obtained from the stationary
solution of the Fokker-Planck equation \eqref{eq:FP}, 
$\partial_t\mathcal{P}_\mathrm{st}=0$, by averaging the position-dependent current 
\begin{align}
\label{eq:I(X)}
\mathcal{I}(X)=&\;\frac{e\Gamma_\mathrm{L}\Gamma_\mathrm{R}}{\Gamma}
\left[
f_\mathrm{F}\left(\frac{F_\mathrm{e} X-e\bar V_\mathrm{g}-eV/2}{k_\mathrm{B}T}\right)
\right.\nonumber\\
&\left.-f_\mathrm{F}\left(\frac{F_\mathrm{e} X-e\bar V_\mathrm{g}+eV/2}{k_\mathrm{B}T}\right)
\right]
\end{align}
with the phase-space distribution, 
\begin{equation}
\label{eq:I}
 I=\iint\mathrm{d}X\mathrm{d}P\,\mathcal{P}_\mathrm{st}(X, P)\mathcal{I}(X).
\end{equation}

Before we proceed, it is convenient to introduce reduced variables in terms of
the relevant energy scale of the problem
$E^0_\mathrm{E}=F_\mathrm{e}^2/m\omega_0^2$, 
the polaronic shift $\ell=F_\mathrm{e}/m\omega_0^2$, and the vibrational frequency for vanishing
compression force $\omega_0$ [see Eq.~\eqref{eq:omega}]. 
By denoting $x=X/\ell$, $p=P/m\omega_0\ell$, $\tau=\omega_0 t$, 
the Fokker-Planck equation \eqref{eq:FP} becomes  
\begin{align}
\label{eq:scaled_FP}
\partial_\tau\mathcal{P}=&
-p\partial_x\mathcal{P}-f_\mathrm{eff}(x)\partial_p\mathcal{P}
+\big(\gamma(x)+\gamma_\mathrm{e}\big)\partial_p(p\mathcal{P})
\nonumber\\
&+\left(\frac{d(x)}{2}+\gamma_\mathrm{e}\tilde T\right)\partial_p^2\mathcal{P}
\end{align}
with the scaled effective force given by
\begin{equation}
\label{eq:f_eff}
  f_\mathrm{eff}(x)=\delta x-\tilde\alpha x^3-n_0(x),
\end{equation}
the reduced force $\delta=F/F_\mathrm{c}-1$, and the anharmonicity parameter $\tilde\alpha=\alpha
\ell^4/E^0_\mathrm{E}$. 
We further introduced a scaled current-induced diffusion constant
\begin{equation}
\label{eq:d}
   d(x)=\frac{2\omega_0}{\Gamma}n_0(x)\left[1-n_0(x)\right],
\end{equation}
and damping coefficient
\begin{equation}
\label{eq:gamma}
    \gamma(x)=-\frac{\omega_0}{\Gamma}\partial_xn_0(x).
\end{equation}
In Eq.~\eqref{eq:scaled_FP}, 
$\gamma_\mathrm{e}=\eta_\mathrm{e}/m\omega_0=Q^{-1}$, where $Q$ is the quality
factor of the mechanical resonator and $\tilde T=k_\mathrm{B}T/E_\mathrm{E}^0$. 
In these scaled units, the electromechanical coupling
appears only in the coefficient of the quartic term, 
$\tilde\alpha=\alpha F_\mathrm{e}^2/(m\omega_0^2)^3$. 
It is important to notice that for actual experiments on suspended carbon 
nanotubes, \cite{hutte09_NL, steel09_Science,lassa09_Science}
$\tilde\alpha \ll 1$ as we will discuss more extensively in 
Sec.~\ref{sect:experiment}.

\section{Mean-field approach: enhancement of the current blockade}
\label{sect:mean-field}
We begin our analysis by assuming 
$\omega_0/\Gamma\rightarrow0$.\cite{footnote:adiabatic}
Note that the diffusion and dissipation coefficients $d(x)$ and $\gamma(x)$ 
in Eq.~\eqref{eq:scaled_FP} are proportional to $\omega_0/\Gamma$ 
[cf.\ Eqs.~\eqref{eq:d} and \eqref{eq:gamma}],  
so that this implies to neglect current-induced fluctuations.
In this limit, the stationary solution for $\mathcal{P}$ is 
given by a Boltzmann distribution at temperature $\tilde T$,
\begin{equation}
\label{eq:Boltzmann}
  \mathcal{P}_\mathrm{st}(x, p)=\mathcal{N}
  \exp{\left(-\frac{p^2/2+v_\mathrm{eff}(x)}{\tilde T}\right)},
\end{equation}
with $\mathcal{N}$ a normalization constant. 
In order to obtain transparent analytical results, we
also assume zero temperature (in fact, $\hbar\omega\ll
k_\mathrm{B}T\ll E_\mathrm{E}^0$),
such that the stationary probability distribution
\eqref{eq:Boltzmann} becomes $\mathcal{P}_\mathrm{st}(x,
p)=\delta(p)\delta(x-x_\mathrm{m})$. Here, $x_\mathrm{m}$ is the global minimum of
the effective potential
\begin{equation}
\label{eq:v_eff_def}
v_\mathrm{eff}(x)=-\int^x\mathrm{d}x'f_\mathrm{eff}(x')
\end{equation}
corresponding to the effective force \eqref{eq:f_eff} and can be determined from
the dynamical equilibrium equation
\begin{equation}
\label{eq:equilibrium}
f_\mathrm{eff}(x)=0, \qquad \frac{\mathrm{d}f_\mathrm{eff}(x)}{\mathrm{d}x}<0.
\end{equation}
Notice that the latter equation can have more than one solution, such that the
system is multi-stable.
In this zero-temperature
limit, the current can then easily be obtained from Eq.~\eqref{eq:I}. 
Doing so as a function of the gate and bias 
voltages, we can determine the Coulomb diamond for a given
compression force $\delta$. 
At zero temperature, one finds that there always exists a region at low bias
voltage where the current is suppressed. 

\begin{figure}[tb]
\includegraphics[width=.89\columnwidth]{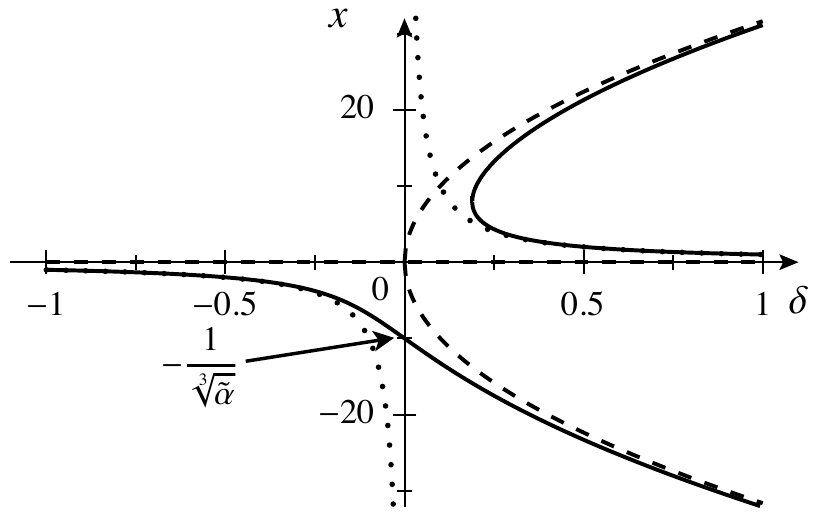}
\caption{%
\label{fig:simple}
Example of a solution $x(\delta)$ of the equation for dynamic equilibrium
\eqref{eq:equilibrium} for $n_0=0$ (dashed line),
$n_0=1$ (solid line), and for $n_0=1$ and $\tilde\alpha=0$ (dotted line). In the
figure, $\tilde\alpha=10^{-3}$.}
\end{figure}

To characterize this classical current blockade, we define 
$\Delta_v$, the minimal value of bias voltage, for which a finite current 
flows through the device at zero temperature. 
It is useful to first derive a simple estimate of the 
maximally obtainable $\Delta_v$.
To do so, we solve the dynamic equilibrium equation \eqref{eq:equilibrium} for $n_0=0$ and 
$n_0=1$, corresponding to empty and occupied central island,
respectively.
For $n_0=0$ one has the solutions $x=0$ for any $\delta$ 
and $x=\pm \sqrt{\delta/\alpha}$ for $\delta>0$ 
of the pristine Euler instability (see Fig.~\ref{fig:simple}, dashed line).
For $n_0=1$, the solutions can easily be sketched for $\tilde\alpha\ll1$  
as an interpolation of the $\tilde\alpha=0$ solutions
(dotted line in Fig.~\ref{fig:simple}) and the solutions for $n_0=0$.
The exact result is shown as a solid line in Fig.~\ref{fig:simple}. 
We are interested in the maximum 
shift in $x$ that the system undergoes in response to a fluctuation of 
$n_0$ by one unit, $\Delta x$.
It is apparent from the figure that this happens for 
$\delta=0$ where $\Delta x=1/\sqrt[3]{\tilde\alpha}$.
The corresponding change in the effective potential \eqref{eq:v_eff_def} is 
$\Delta v_\mathrm{eff}\sim1/\sqrt[3]{\tilde\alpha}$. 
This provides an estimate of the maximal energy gap generated by 
the electromechanical coupling, and thus a good estimate
of $\Delta_v$. 
Notice that the simple argument above is not specific to the transport model we
are considering here, as the specific form of $n_0$ in the conducting region
does not enter our argument. Thus, we expect that our estimate of a
maximal gap $\Delta_v\sim1/\sqrt[3]{\tilde\alpha}$ remains valid for metallic
quantum dots, as well as in the resonant transport regime.

We now turn to the complete solution of Eq.~\eqref{eq:equilibrium}. For
simplicity, 
we assume symmetric coupling to the leads
($\Gamma_\mathrm{L}=\Gamma_\mathrm{R}=\Gamma/2$), such that 
the average occupation of the dot at fixed $x$ [entering into the effective force
\eqref{eq:f_eff}] is obtained from the zero-temperature limit of 
Eqs.~\eqref{eq:Gamma01} and \eqref{n0Gen} and given by
\begin{equation}
	\label{eq:n0}
  n_0(x)=\frac 12\left[
  \Theta\left(-x+v_\mathrm{g}+\frac{v}{2}\right)+\Theta\left(-x+v_\mathrm{g}-\frac{v}{2}\right)
  \right],
\end{equation}
where $v=eV/E^0_\mathrm{E}$ (assumed positive for definiteness), $v_\mathrm{g}=e\bar
V_\mathrm{g}/E^0_\mathrm{E}$, and $\Theta(z)$ is the Heaviside step function. 
In the most general case, we solve Eq.~\eqref{eq:equilibrium} numerically. 
However, transparent analytical expressions can be obtained in the limits 
$|\delta|\gg\sqrt[3]{\tilde\alpha}$ and $|\delta| \ll \sqrt[3]{\tilde\alpha}$.
In particular, we
can obtain explicit expressions for the value of $v$ beyond which the 
current begins to flow, i.e., the gap $\Delta_v$.
To first order in the small parameter $\sqrt[3]{\tilde\alpha}/|\delta|$ (far
from the instability) and $|\delta|/\sqrt[3]{\tilde\alpha}$ (in the vicinity of
the instability), we find that 
\begin{equation}
\label{eq:gap}
\Delta_v=
\begin{cases}
\displaystyle
\vspace{.2truecm}
-\frac{1}{2\delta}, & -1\leqslant\delta\ll-\sqrt[3]{\tilde\alpha},\\
\vspace{.2truecm}
\displaystyle
\frac{1}{4\delta}, & \delta\gg\sqrt[3]{\tilde\alpha},\\
\displaystyle
\frac{\sqrt[3]{2}-1}{\sqrt[3]{\tilde\alpha}}
\left(\frac{3}{2^{4/3}}-\frac{\delta}{\sqrt[3]{\tilde\alpha}}\right), 
& |\delta|\ll\sqrt[3]{\tilde\alpha}.
\end{cases}
\end{equation}
Far from the mechanical instability, the gap is simply given by the result of
a harmonic theory (see Appendix~\ref{sec:harmonic}) where $\Delta_v=1/2v''(\bar x)$, with $v''(\bar x)$ the
curvature of the bare potential $v(x)$ [i.e., the effective potential without the
contribution from $n_0(x)$] at its global minimum $\bar x$.
Far below the buckling instability 
($-1\leqslant\delta\ll-\sqrt[3]{\tilde\alpha}$), $\bar x=0$ and $v''(\bar
x)=-\delta$, such
that $\Delta_v=-1/2\delta$. Far above the mechanical instability
($\delta\gg\sqrt[3]{\tilde\alpha}$),
$\bar x=-\sqrt{\delta/\tilde\alpha}$ and $v''(\bar x)=2\delta$, such that 
$\Delta_v=1/4\delta$. As one approaches the buckling instability
from below or above,
the apparent divergences in the first two lines of Eq.~\eqref{eq:gap} are cutoff 
by the cubic term in $x$ in the
effective force \eqref{eq:f_eff}, and for $|\delta|\ll\sqrt[3]{\tilde\alpha}$,
the maximal gap $\Delta_v\sim1/\sqrt[3]{\tilde\alpha}$ is reached. 

\begin{figure}[tb]
\includegraphics[width=\columnwidth]{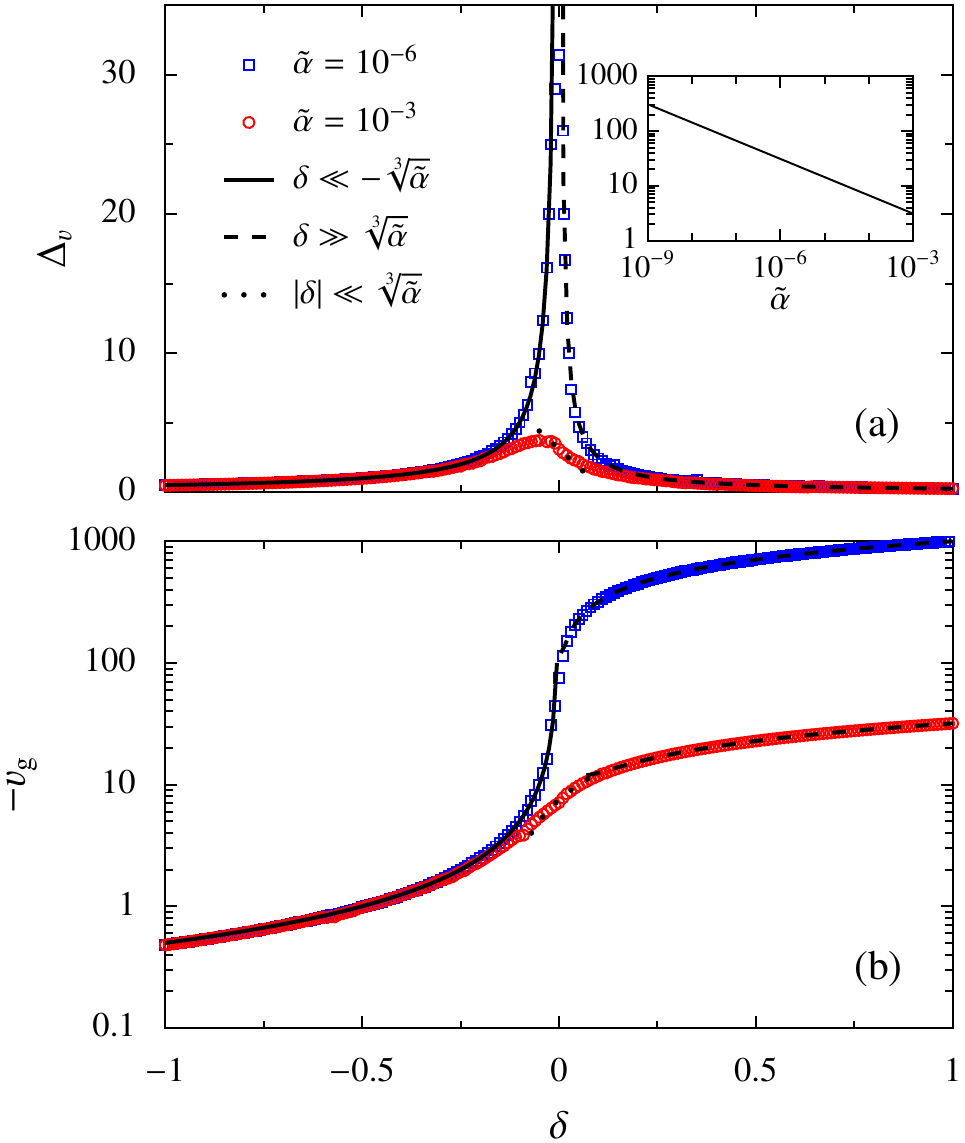}
\caption{\label{fig:gap_mean_field}%
(Color online) (a) Gap $\Delta_v$ and (b) gate voltage $v_\mathrm{g}$ 
(defined as the bias and gate voltages in reduced units at
the apex of the Coulomb diamond, respectively) as a function of
the scaled compression force $\delta=F/F_\mathrm{c}-1$. The red circles and blue squares
are numerical results for $\tilde\alpha=10^{-3}$ and $10^{-6}$, respectively, which are
compared to the asymptotic behaviors \eqref{eq:gap} and \eqref{eq:vg} 
for forces below (solid line), above
(dashed line), and in the vicinity (dotted line) of 
the critical force $F_\mathrm{c}$. 
Inset: Gap $\Delta_v\sim1/\sqrt[3]{\tilde\alpha}$ 
from Eq.~\eqref{eq:gap} as a function of $\tilde\alpha$ at
the mechanical instability ($\delta=0$).}
\end{figure}

The analytical results of Eq.~\eqref{eq:gap} are compared to a numerical 
calculation of the gap in Fig.~\ref{fig:gap_mean_field}(a) for
$\tilde\alpha=10^{-3}$
and $\tilde\alpha=10^{-6}$ (red dots and blue squares in the figure, respectively). 
It is evident from the figure that there is a
dramatic increase of the gap close to the instability.
Furthermore, the smaller $\tilde\alpha$, i.e., the smaller the electromechanical coupling, 
the larger is the increase of the gap at the instability 
relative to its value for vanishing compression force [see the inset in
Fig.~\ref{fig:gap_mean_field}(a)]. However, of course, the maximal value of the
gap in absolute terms increases with the strength of the electromechanical coupling as
$F_\mathrm{e}^{4/3}$.
It would thus be of great experimental interest to exploit the Euler instability to obtain a clear
signature of the classical current blockade in transport experiments on suspended quantum
dots. 

\begin{figure*}[bth]
\includegraphics[width=1.8\columnwidth]{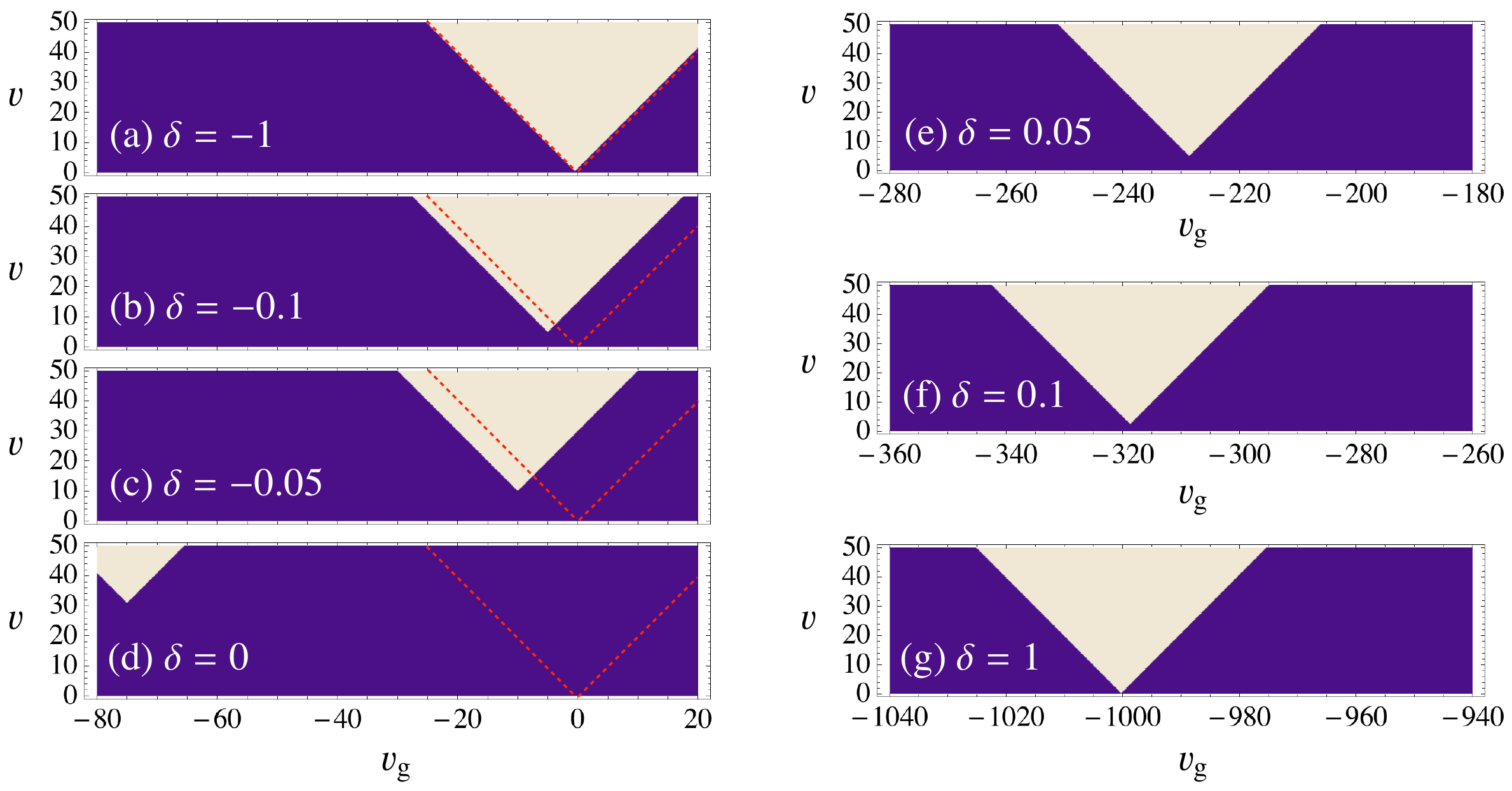}
\caption{\label{fig:CB_diamond}%
(Color online) Mean-field current $I$ at zero temperature and for symmetric
coupling to the leads ($\Gamma_\mathrm{L}=\Gamma_\mathrm{R}=\Gamma/2$), as a function of bias
$v$ and gate voltage $v_\mathrm{g}$ 
(measured in units of the elastic energy $E_\mathrm{E}^0$).
The (scaled) compression force $\delta$ increases from (a) to (g).
Notice that the scale of the $v_\mathrm{g}$-axis is different in
(e), (f), and (g), and in (a)-(d). 
The red dashed lines indicate the position of the
Coulomb diamond in the absence of electromechanical coupling ($F_\mathrm{e}=0$). In the
figure, $\tilde\alpha=10^{-6}$, and 
dark blue and white regions correspond to $I=0$ and $I=e\Gamma/4$, respectively.}
\end{figure*}

The gaps of Eq.~\eqref{eq:gap} are obtained for values of the gate voltage
approximately given by
\begin{equation}
\label{eq:vg}
v_\mathrm{g}=
\begin{cases}
\displaystyle
\vspace{.2truecm}
\frac{1}{2\delta}, & -1\leqslant\delta\ll-\sqrt[3]{\tilde\alpha},\\
\vspace{.2truecm}
\displaystyle
-\frac{1}{4\delta}-\sqrt{\frac{\delta}{\tilde\alpha}}, & \delta\gg\sqrt[3]{\tilde\alpha},\\
\displaystyle
-\frac{1}{4\sqrt[3]{\tilde\alpha}}\left(3+\frac{2\delta}{\sqrt[3]{\tilde\alpha}}\right), 
& |\delta|\ll\sqrt[3]{\tilde\alpha},
\end{cases}
\end{equation}
which are shown in Fig.~\ref{fig:gap_mean_field}(b) and compared to
a numerical calculation. Equations~\eqref{eq:gap} and \eqref{eq:vg} 
define the apexes of the Coulomb diamonds which are shown in
Fig.~\ref{fig:CB_diamond}.
The effect of the compression force is thus to continuously displace the Coulomb
diamond in the $v$-$v_\mathrm{g}$ plane towards negative gate voltages [see also
Fig.~\ref{fig:gap_mean_field}(b)], and to open a gap which is maximal close to
the Euler instability at $\delta=0$ [see Fig.~\ref{fig:CB_diamond}(d)].
Note that the shift in gate voltage is strongly asymmetric about the Euler
instability. While the shifts are only small below the Euler instability [see
Figs.~\ref{fig:CB_diamond}(a)-(c) and Fig.~\ref{fig:gap_mean_field}(b)], the shifts in gate voltage are orders of magnitudes
larger on the buckled side of the Euler instability [see
Figs.~\ref{fig:CB_diamond}(e)-(g) and Fig.~\ref{fig:gap_mean_field}(b)]. In fact, it may be that
these shifts would be the most easily detected consequence of the Euler buckling
instability in NEMS. 
In Fig.~\ref{fig:CB_diamond}, the bias and gate voltages are measured in units of the elastic
energy $E_\mathrm{E}^0$ which is of the order of a few $\mu$eV for typical
experiments on suspended carbon nanotubes (see Sec.~\ref{sect:experiment}).
The smallness of this energy scale explains why the scaled numerical values of
the shifts become so large on the buckled side of the Euler instability.

It is also interesting to comment on the shape of the Coulomb
blockade diamond. In Ref.~\onlinecite{weick10_PRB}, we showed for the case of intrinsic
electron-phonon coupling (quadratic in $x$) and for a metallic quantum dot 
that the Euler buckling instability
leads to nonlinear deformations of the Coulomb diamonds, a phenomenon that we
have named ``tricritical current blockade". In contrast, our
present results show that for a capacitive electromechanical coupling (linear in
$x$) and for a single-level quantum dot, 
the shape of the Coulomb diamond remains unchanged. The conventional
triangular shape in the $v$-$v_\mathrm{g}$ plane is delineated by straight lines with
$v\sim\pm 2v_\mathrm{g}$ for any value of the compressive strain. As we have checked,
\cite{weick_unpublished} the difference between the present results and those of
Ref.~\onlinecite{weick10_PRB} is due to the difference in the transport models
considered (metallic vs.\ single-level quantum dot), and not to the type of
electromechanical coupling (intrinsic vs.\ extrinsic). Specifically, we find that
the difference is due to the fact that in the single-level case, the average occupation
of the dot abruptly jumps as a function of gate voltage, while in the metallic
case, this occupation gradually changes due to the continuous density of states
of the dot.
Notice also that for
intrinsic electromechanical coupling, the Coulomb diamond is not influenced by the
latter for compression forces below the critical force ($\delta<0$), in contrast
to the present case. This is due to the fact that, in the flat state, the quadratic 
electromechanical coupling merely represents a renormalization of the
fundamental bending mode
frequency, and does not lead to current blockade. 

It is instructive to make the analogies with standard results of Landau mean
field theory for continuous phase transitions \cite{chaikin} explicit. According to
Eq.~\eqref{eq:equilibrium}
governing the dynamical equilibrium, we can make the following identifications:
$x$ corresponds to the order parameter in Landau theory, $\delta$ to the reduced
temperature, and $\tilde\alpha$ to the coefficient of the quartic term in the
Landau free energy. Finally, $n_0$ plays a role similar to a symmetry-breaking
(magnetic) field. In the present case, this field is in general dependent on $x$,
which has no correspondence in Landau theory. Nevertheless, the analogy between
$n_0$ and a magnetic field is helpful since some of our results can be understood
by comparing the situations with zero ($n_0=0$) and one ($n_0=1$) electrons on
the dot, as illustrated by the above estimate for the maximal $\Delta_v$ (see
Fig.~\ref{fig:simple}).   

With these correspondences, we can now establish analogies between some of our
results and standard results of Landau theory. To start with, the dependence of
the displacement $x\sim\pm\delta^{1/2}$ in the buckled state is analogous to the result of Landau
theory that the order parameter exponent is $\beta=1/2$. 
Given that $\Delta_v$ depends linearly on $x$, we can also interpret the
relations in Eq.~\eqref{eq:gap} in terms of Landau theory. Let us start with the case of
small $\delta$, in the immediate vicinity of the instability. In this case,
we find that $\Delta_v\sim \tilde\alpha^{-1/3}$. The exponent of $\tilde\alpha$ corresponds to
the critical exponent $\delta=3$ of Landau theory governing the dependence of the order parameter on
the symmetry-breaking field at the critical temperature. 
Further from the instability, we have $\Delta_v\sim 1/|\delta|$. This relation
is related to the familiar Curie law for the order parameter (or the
susceptibility) as function of temperature in an external field, with mean-field
critical exponent $\gamma=1$.

Let us finally emphasize that the analogy with Landau theory is restricted to
mean-field level, since contrarily to critical phenomena where an infinite
number of modes is present, the system we describe is constituted by a single
mode. Moreover, beyond mean-field theory, fluctuations in Landau theory are
purely thermal, while in the present context, non-equilibrium fluctuations play
an essential role. It is the effects of these fluctuations which we turn to in
the next section.

\section{Thermal and current-induced fluctuations}
\label{sect:temperature}
We now go beyond the mean-field results of the previous section by taking
into account the effects of the thermal as well as the current-induced
fluctuations. 
It is physically clear that these fluctuations will lead
to a smoothening of the current blockade at low bias voltages, as the system can
explore more conducting states in phase space.

\subsection{Temperature effects}
\label{sec:temp}
\begin{figure}[tb]
\includegraphics[width=\columnwidth]{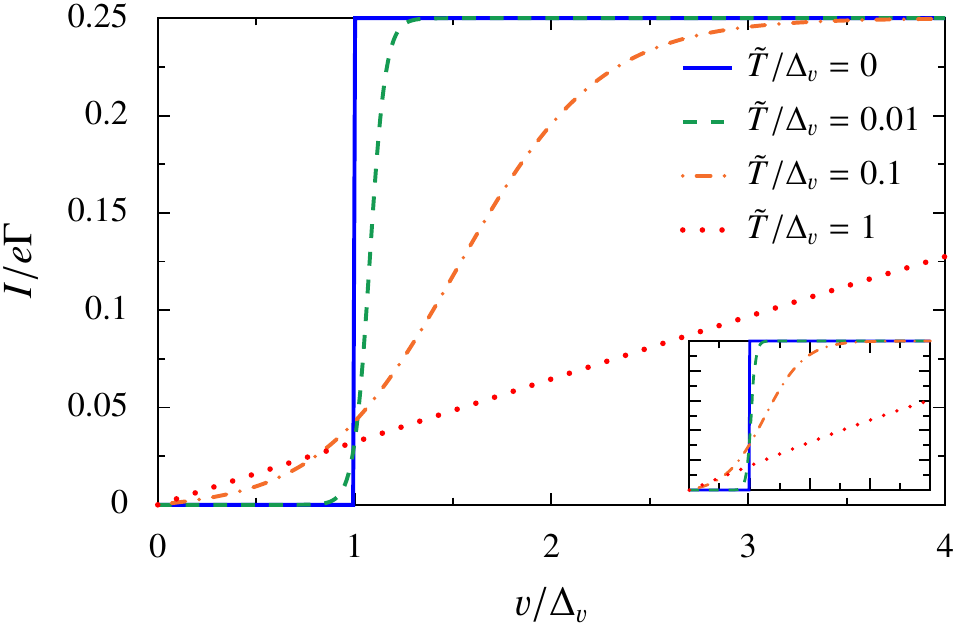}
\caption{
\label{fig:extrinsic_dissipation}%
(Color online) 
Current $I$
at the apex of the Coulomb diamond as a function of bias $v$ scaled
by the energy gap $\Delta_v$ for various values of $\tilde T/\Delta_v$,
and for 
compression forces in the vicinity of the buckling instability
($|\delta|\ll\sqrt[3]{\tilde\alpha}$). In the figure, only the
temperature-induced fluctuations are considered. 
Inset: Same as the main figure for compression forces far from the buckling
instability ($|\delta|\gg\sqrt[3]{\tilde\alpha}$).}
\end{figure}

We first neglect the current-induced fluctuations and focus on thermal
fluctuations only. As discussed in Sec.~\ref{sect:mean-field}, 
this becomes asymptotically exact in the 
extreme adiabatic limit of $\omega_0/\Gamma\rightarrow0$, where
the terms $\gamma(x)$ and $d(x)$ can be dropped 
from Eq.~\eqref{eq:scaled_FP}. 
The stationary solution for $\mathcal{P}$ is then
given by the Boltzmann distribution \eqref{eq:Boltzmann}
with the effective potential
\begin{align}
\label{eq:v_eff}
    v_\mathrm{eff}(x)=&\,-\frac{\delta x^2}{2}+\frac{\tilde\alpha x^4}{4}+x
    +\frac{\tilde T}{2}
   \ln{\Bigg(
   f_\mathrm{F}\left(\frac{x-v_\mathrm{g}+v/2}{\tilde T}\right)\Bigg)}
   \nonumber\\
    &\,+\frac{\tilde T}{2}
   \ln{\Bigg(
   f_\mathrm{F}\left(\frac{x-v_\mathrm{g}-v/2}{\tilde T}\right)\Bigg)}.
\end{align}
The current can now be easily calculated by numerical integration of 
Eq.~\eqref{eq:I} with Eq.~\eqref{eq:Boltzmann}.
The result is shown in Fig.~\ref{fig:extrinsic_dissipation} as a function of
the bias voltage, for gate voltages corresponding to the apex of the 
modified zero-temperature Coulomb diamond [cf.\ Eq.~\eqref{eq:vg}].
Once plotted as a function of $v/\Delta_v$, one finds that the current behavior 
is similar at the transition 
(Fig.~\ref{fig:extrinsic_dissipation}) and far from the transition (inset of
Fig.~\ref{fig:extrinsic_dissipation}).
In both cases, the low-bias blockade of
the current becomes less pronounced as 
temperature increases, and vanishes completely for temperatures of the order of the gap
$\Delta_v$. As shown in Appendix~\ref{app:temp} [cf.\ Eq.~\eqref{eq:I_lowT}], the current has a
Fermi-function-like behavior as a function of the bias voltage
for temperatures much smaller than the energy gap $\Delta_v$ 
(see dashed and dashed-dotted lines in Fig.~\ref{fig:extrinsic_dissipation}). 
It is thus exponentially suppressed for bias voltage below the gap. At larger
temperatures, Eq.~\eqref{eq:I_highT} shows that the current
is linear in the bias voltage (see dotted line in Fig.~\ref{fig:extrinsic_dissipation}).

Our numerical and analytical results (cf.\ Appendix~\ref{app:temp}) thus confirm 
that tuning the system near the buckling instability where
$\Delta_v$ dramatically increases allows one
to enlarge the temperature region over which the current blockade is 
observable.

\subsection{Non-equilibrium dynamics close to the mechanical instability}
We now consider the non-equilibrium Langevin dynamics of the nanobeam by solving the full 
Fokker-Planck equation \eqref{eq:scaled_FP}. 
This is done by discretization of the 
Fokker-Planck equation and solution of the resulting 
linear system.
We focus on the transition region ($\delta=0$) and calculate the current for 
$v_\mathrm{g}$ at the apex of the Coulomb diamond [see Eq.~\eqref{eq:vg} and
Fig.~\ref{fig:gap_mean_field}(b)] and temperature
lower than the gap $\tilde T/\Delta_v=0.1$. Before we present our results, we
notice that for $(\omega_0/\Gamma, \gamma_\mathrm{e})\ll1$, we can show that the stationary
distribution of the Fokker-Planck equation approximately only depends on the ratio
$\frac{\gamma_\mathrm{e}}{\omega_0/\Gamma}$, a result we have also checked
numerically (see Appendix~\ref{sec:scaling} for details).
The reason for this behavior is that, for
$(\omega_0/\Gamma, \gamma_\mathrm{e})\ll1$, the stationary distribution is almost a function of the
(reduced) energy $E=p^2/2+v_\mathrm{eff}(x)$ only.

\begin{figure}[tb]
\includegraphics[width=\columnwidth]{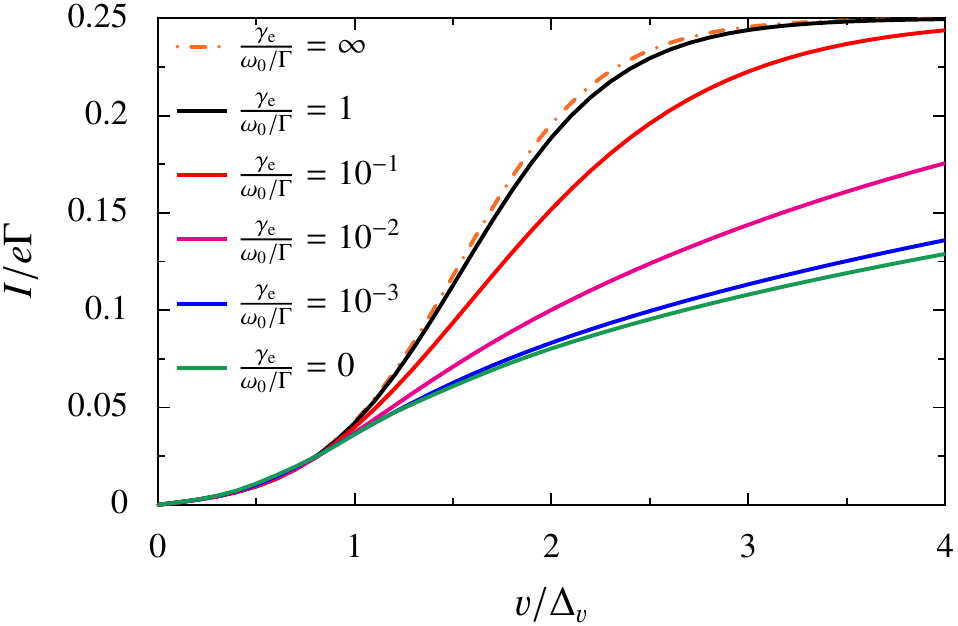}
\caption{\label{fig:fluctuations}%
(Color online) 
Current $I$ at the apex of the Coulomb diamond for $\delta=0$ as a function of
$v/\Delta_v$ for $\tilde\alpha=10^{-6}$ and $\tilde T/\Delta_v=0.1$. Solid
lines: non-equilibrium Langevin dynamics for
$\frac{\gamma_\mathrm{e}}{\omega_0/\Gamma}=1$, $10^{-1}$, $10^{-2}$, $10^{-3}$, and 0, from the
highest to the lowest curve at large bias. Dashed-dotted line: fully adiabatic
limit ($\frac{\gamma_\mathrm{e}}{\omega_0/\Gamma}=\infty$), i.e., 
current only including thermal fluctuations (cf.\ dashed-dotted curve
in Fig.~\ref{fig:extrinsic_dissipation}). In our numerical calculations, we used
$\omega_0/\Gamma=10^{-2}$.}
\end{figure}

Numerical results for the current are shown in Fig.~\ref{fig:fluctuations} 
for various ratios of the inverse quality factor
$Q^{-1}=\gamma_\mathrm{e}$ as quantified by the damping coefficient
$\gamma_\mathrm{e}$ and the adiabaticity parameter $\omega_0/\Gamma$.
Our principal observation is that the current blockade becomes sharper
for low-$Q$ resonators.

One can qualitatively understand the behavior of the current in
Fig.~\ref{fig:fluctuations} by defining an
effective temperature of the system
\begin{equation}
\label{eq:T_eff}
\tilde T_\mathrm{eff}=\frac{\langle d\rangle/2+\gamma_\mathrm{e}\tilde T}
{\langle\gamma\rangle+\gamma_\mathrm{e}}
\end{equation}
in close analogy with the fluctuation-dissipation theorem. \cite{footnote:Teff}
In Eq.~\eqref{eq:T_eff}, 
$\langle d\rangle$ and $\langle\gamma\rangle$ are the averages over the 
phase-space probability distribution of the current-induced fluctuations and dissipation [cf.\
Eqs.~\eqref{eq:d} and \eqref{eq:gamma}], respectively. Notice that the strength
of these two quantities is controlled by the adiabaticity parameter
$\omega_0/\Gamma$. 

\begin{figure}[tb]
\includegraphics[width=\columnwidth]{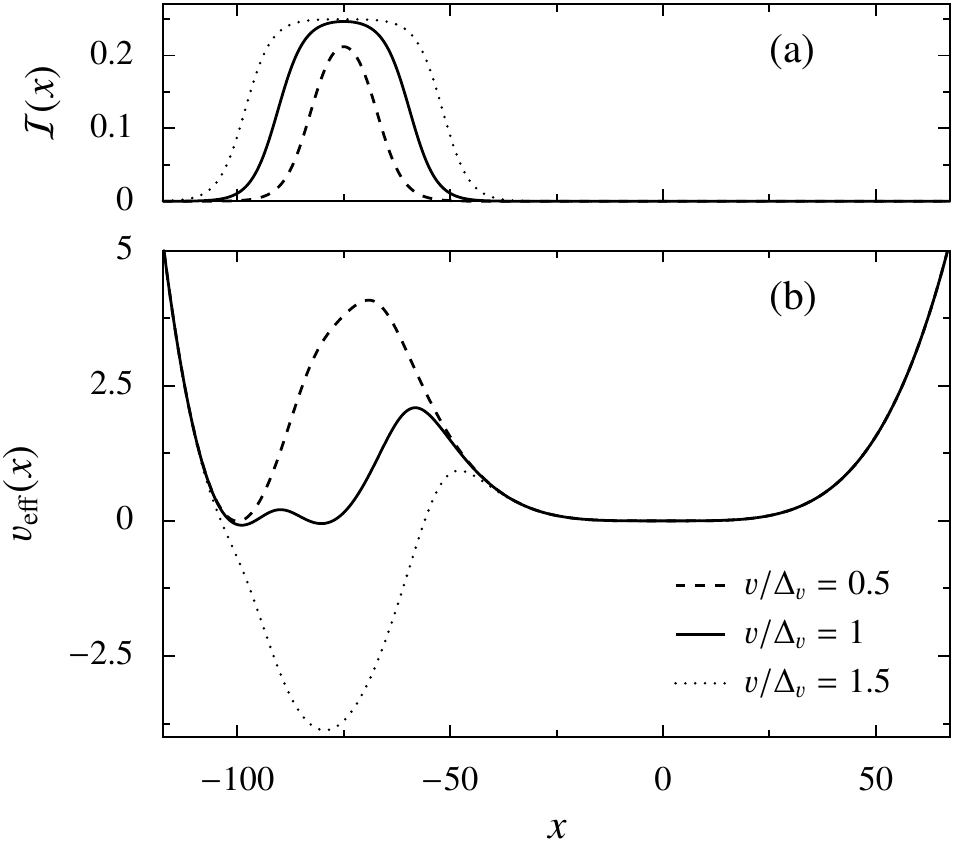}
\caption{\label{fig:v_eff}%
(a) Current $\mathcal{I}$ as a function of $x$ [Eq.~\eqref{eq:I(X)}] and (b)
effective potential $v_\mathrm{eff}(x)$ [Eq.~\eqref{eq:v_eff}] for bias voltages
below (dashed line), above (dotted line), and at (solid line) the energy gap
$\Delta_v$. 
The parameters are the same as in Fig.~\ref{fig:fluctuations}, i.e., $\delta=0$,
$\tilde\alpha=10^{-6}$, $\tilde T/\Delta_v=0.1$, and
$v_\mathrm{g}=-3/4\sqrt[3]{\tilde\alpha}$.}
\end{figure}

As one can see from Fig.~\ref{fig:fluctuations}, for $v<\Delta_v$, 
the current is almost insensitive to the
quality factor, and is the same as without current-induced fluctuations (see
dashed-dotted line in Figs.~\ref{fig:extrinsic_dissipation} and
\ref{fig:fluctuations}). Using Eqs.~\eqref{eq:d} and \eqref{eq:gamma}, we have
\begin{equation}
d(x)=2\gamma(x)\tilde T+\frac{\omega_0}{2\Gamma}
\left[f_\mathrm{F}\left(\frac{x-v_\mathrm{g}-v/2}{\tilde T}\right)
-f_\mathrm{F}\left(\frac{x-v_\mathrm{g}+v/2}{\tilde T}\right)\right]^2
\end{equation}
for symmetric coupling to the leads.
However, for $v<\Delta_v$,
positions $x$ for which the current $\mathcal{I}(x)$ of
Eq.~\eqref{eq:I(X)} is suppressed are most stable (see dashed line in
Fig.~\ref{fig:v_eff}), such
that the current-induced diffusion and damping constants approximately satisfy a ``local"
fluctuation-dissipation theorem for all relevant positions $x$ that are
significantly populated, $d(x)\simeq2 \tilde T\gamma(x)$. 
We thus have
$\langle d\rangle\simeq2\tilde T\langle\gamma\rangle$, and according to the
definition \eqref{eq:T_eff}, we have $\tilde T_\mathrm{eff}\simeq\tilde T$.
Hence, for bias voltages lower than the energy gap $\Delta_v$, the
current-induced fluctuations behave as the thermal ones, essentially keeping the
mechanical system at equilibrium.

On the contrary, for $v>\Delta_v$, positions $x$ for which the system is
conducting are the most stable ones (see dotted line in Fig.~\ref{fig:v_eff}), 
and one has $\langle
d\rangle\simeq\omega_0/2\Gamma$, while $\langle\gamma\rangle$ is exponentially
small. The mechanical system is then subject to strong non-equilibrium
fluctuations. For $\langle\gamma\rangle\ll\gamma_\mathrm{e}$,
we thus have from Eq.~\eqref{eq:T_eff} $\tilde T_\mathrm{eff}\simeq\tilde
T+\frac{\omega_0/\Gamma}{4\gamma_\mathrm{e}}$. This estimate of the effective temperature
shows that the system becomes ``hotter" as the ratio
$\frac{\gamma_\mathrm{e}}{\omega_0/\Gamma}$ decreases. \cite{footnote:hot} Hence, the system can explore
more states in phase space for which $\mathcal{I}(x)$ is suppressed, and, in turn, the current
decreases for decreasing $\frac{\gamma_\mathrm{e}}{\omega_0/\Gamma}$ for $v>\Delta_v$ (see
Fig.~\ref{fig:fluctuations}). The latter argument breaks down when
$\gamma_\mathrm{e}\ll\langle\gamma\rangle$. In that case, we can estimate the
effective temperature self-consistently, by assuming that the phase-space
distribution is a Boltzmann distribution at the temperature $\tilde
T_\mathrm{eff}$. Approximating the effective potential by its zero temperature
expression, and averaging $d(x)$ and $\gamma(x)$ over the effective Boltzmann
distribution, we find for $v\gg\Delta_v$
\begin{equation}
\tilde T_\mathrm{eff}=\frac{\pi\Delta_v}{128A}
\exp{\left(\frac{Av^2}{\Delta_v\tilde T_\mathrm{eff}}\right)}, 
\end{equation}
with $A=9(1-2^{-1/3})/2^{11/3}$.
We thus have $\tilde T_\mathrm{eff}/\Delta_v\sim
(v/\Delta_v)^2/\ln{(v/\Delta_v)}\gg\tilde T/\Delta_v$, which explains
why for $\gamma_\mathrm{e}=0$ the current is more suppressed than for finite
$\gamma_\mathrm{e}$. It is also interesting to note that this estimate of the
effective temperature is much larger than for a metallic quantum dot, where
$\tilde T_\mathrm{eff}\sim v$ (Ref.~\onlinecite{armou04_PRB}). The reason for this difference
is that, in the metallic case, the fluctuation and the dissipation are of the
same order inside the bias window, while in the single-level case, the
average dissipation is exponentially suppressed as $\gamma(x)$ only has a significant contribution
for positions $x$ corresponding to the borders of the Coulomb diamond [see
Eq.~\eqref{eq:gamma}]. 

Our results show that a low quality factor is 
more suitable for the observation of the current blockade 
in classical resonators. 
It is interesting to note that this conclusion is also
valid in the
quantum case, \cite{koch05_PRL, koch06_PRB} where the Franck-Condon blockade is
more pronounced for fast equilibration of the vibron mode.
Due to the scaling of our results for the classical current blockade with the parameter
$\frac{\gamma_\mathrm{e}}{\omega_0/\Gamma}$ (see Fig.~\ref{fig:fluctuations}),
we also conclude that it is
advantageous for the observation of this phenomenon to have a
resonator which is slow compared to the tunneling dynamics, i.e., 
$\omega_0\ll\Gamma$.

\section{Effect of a finite excess charge on the quantum dot}
\label{sec:average_charge}
Within the transport model of a single resonant
electronic level with infinite charging energy that we have used so far, the
number of electrons
on the dot can only vary between 0 or 1 (see Sec.~\ref{Model}). More generally,
the range of gate voltages can 
exceeds the charging energy and 
the average number of excess electrons $N$ on the dot can be much larger than 1. 
Due to these excess electrons, an additional
force $-F_N$ further bends the nanotube, and hence increases its
vibrational frequency. \cite{steel09_Science, lassa09_Science} 
We can thus expect that the bias voltage below which the current is blocked will
decrease when $N$ increases. 

\begin{figure}[tb]
\includegraphics[width=\columnwidth]{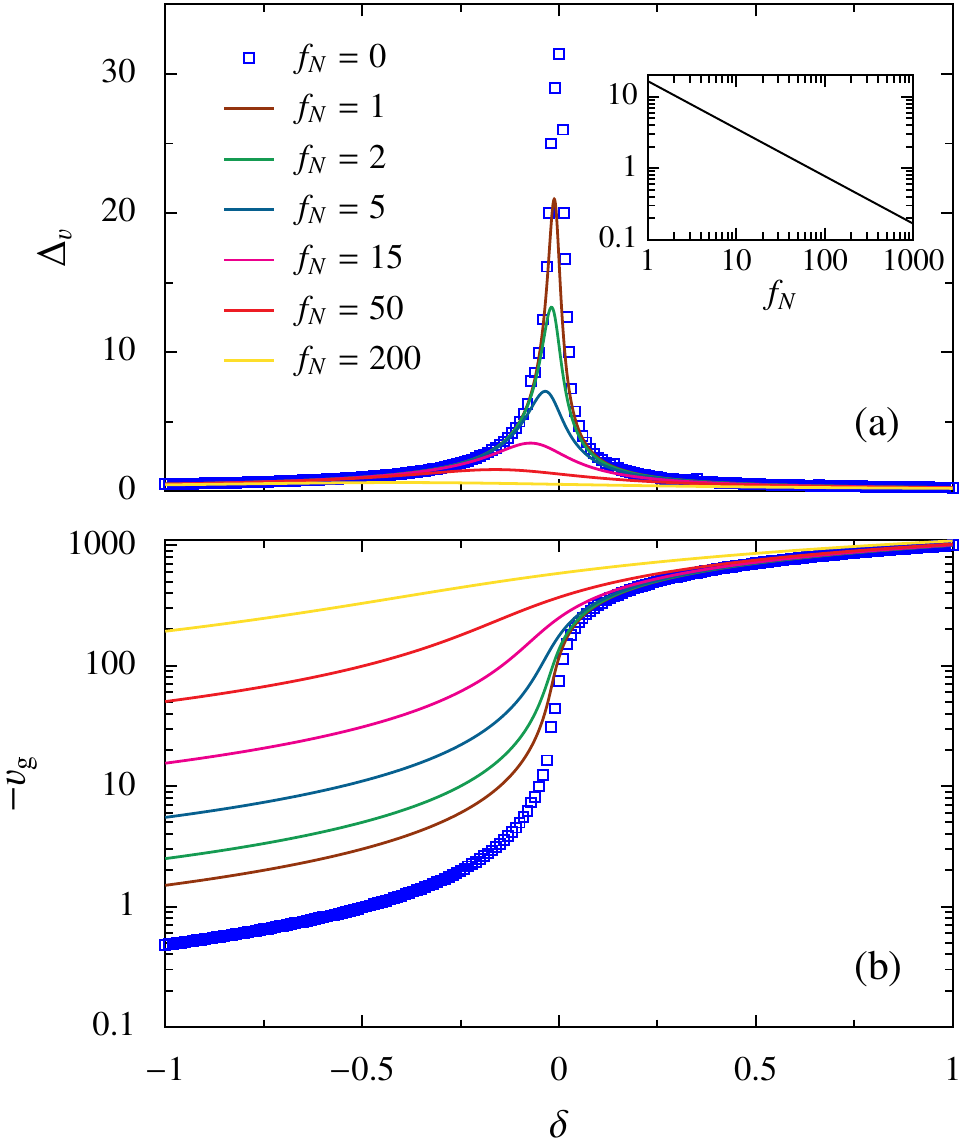}
\caption{\label{fig:offset_charge}%
(Color online) (a) Gap $\Delta_v$ and (b) gate voltage $v_\mathrm{g}$ at the
apex of the Coulomb diamond for
$\tilde\alpha=10^{-6}$
as a function of
the scaled compression force $\delta$ for increasing values of $f_N$ [from top
to bottom and bottom to top at $\delta=0$ in (a) and (b), respectively]. 
The blue squares ($f_N=0$) corresponds to the 
numerical results of Fig.~\ref{fig:gap_mean_field}, while the solid lines result
from the harmonic approximation (see text).
Inset: Gap $\Delta_v\sim1/f_N^{2/3}$ 
from Eq.~\eqref{eq:gap_fN} as a function of $f_N$ at
the mechanical instability ($\delta=0$).}
\end{figure}

In order to investigate the effect of a non-vanishing average excess charge on the
quantum dot, we assume that the gate voltage is such that there is
either $N$ or $N+1$ electrons on the dot. 
We measure the fluctuation of the dot occupation $n_\mathrm{d}$ with respect to $N$, and 
incorporate the resulting additional force in
Eq.~\eqref{eq:f_eff} by writing $f_\mathrm{eff}(x)=\delta x-\tilde\alpha
x^3-n_0(x)-f_N$, where $f_N=F_N/F_\mathrm{e}$. 
We neglect thermal and current-induced fluctuations, and work within a
mean-field approximation at zero temperature, such that $n_0(x)$
is given by Eq.~\eqref{eq:n0}. For finite $f_N$, the bare potential $v(x)$ (i.e.,
without the current-induced contribution) can be approximated by a harmonic
potential close to its global minimum $\bar x$, such that the bias voltage below
which the current is blocked is given by $\Delta_v=1/2v''(\bar x)$ (see
Appendix~\ref{sec:harmonic}). The energy gap (resulting from the most stable
solution of $\delta \bar x-\tilde\alpha\bar x^3=f_N$) is plotted in
Fig.~\ref{fig:offset_charge}(a), and the gate voltage at the apex of the Coulomb
diamond in Fig.~\ref{fig:offset_charge}(b).
As anticipated, the increase of the energy gap close to the mechanical
instability is reduced as $f_N$ increases. Moreover, the displacement of the
Coulomb diamond in $v_\mathrm{g}$ is less pronounced for large $f_N$. 

Far from ($|\delta|\gg\sqrt[3]{\tilde\alpha f_N^2}$) and in the vicinity of
($|\delta|\ll\sqrt[3]{\tilde\alpha f_N^2}$) the Euler instability, we
analytically find for the gap
\begin{equation}
\label{eq:gap_fN}
\Delta_v=
\begin{cases}
\displaystyle
\vspace{.2truecm}
-\frac{1}{2\delta}, & -1\leqslant\delta\ll-\sqrt[3]{\tilde\alpha f_N^2},\\
\displaystyle
\frac{1}{4\delta}, & \delta\gg\sqrt[3]{\tilde\alpha f_N^2},\\
\displaystyle
\frac{1}{6\sqrt[3]{\tilde\alpha
f_N^2}}\left(1-\frac{\delta}{3\sqrt[3]{\tilde\alpha f_N^2}}\right),
& |\delta|\ll\sqrt[3]{\tilde\alpha f_N^2}, 
\end{cases}
\end{equation}
to first order in the small parameter $\sqrt[3]{\tilde \alpha f_N^2}/|\delta|$
(far from the instability) and $|\delta|/\sqrt[3]{\tilde \alpha f_N^2}$ (in the
vicinity of the instability).
Far below and above the instability, the gap follows the same behavior as for
$f_N=0$ [see Eq.~\eqref{eq:gap}]. This is due to the fact that for large
$|\delta|\gg f_N$, the stable position of the beam is similar to the one for
$f_N=0$. In the vicinity of the instability, the gap is reduced as $f_N$
increases as $1/f_N^{2/3}$
(see the inset in Fig.~\ref{fig:offset_charge}). The reduction of the maximal gap close
to the instability is a direct consequence of the smoothening of the mechanical
transition between flat and buckled states due to the presence of the
symmetry-breaking force $f_N$, similar to the behavior of the order parameter
at a second-order phase transition in a symmetry-breaking field. \cite{chaikin}

We can estimate the force above which the increase of the gap at the instability
completely vanishes by equating in Eq.~\eqref{eq:gap_fN} the gap at, say,
$\delta=-1$ and $\delta=0$. We obtain that the increase of the gap should vanish
once $f_N\gtrsim1/\sqrt{3^3\tilde\alpha}$. Since for large $N$, $F_N\simeq
F_\mathrm{e}N/2$, this means that if the average charge on the dot
$N\gtrsim2/\sqrt{3^3\tilde\alpha}$, the increase of the gap at the instability
completely disappears. Since $\tilde\alpha$ is typically small 
(see Sec.~\ref{sect:experiment}), we expect that a significant
increase of the current blockade at the mechanical instability persists
for a wide range of gate voltages.

\section{Experimental realization}
\label{sect:experiment}
The electromechanical coupling \eqref{eq:Hc} is typically weak in experiment.
For this reason, only a precursor of the classical current blockade has been
seen in two recent experiments on suspended carbon nanotube quantum dots, \cite{steel09_Science, lassa09_Science} 
but the full current blockade has not yet been observed.  
Indeed, we can obtain an estimate for the frequency shift of the fundamental
bending mode,
induced by the electromechanical coupling, from the effective potential associated
with $F_\mathrm{eff}(X)$. The shift arises from the position dependence of $n_0(X)$.
Expanding the current-induced force for weak electromechanical coupling, we find
\begin{equation}
  F_\textrm{c-i}(X)\simeq-F_\mathrm{e} n_0(0)
  -F_\mathrm{e}^2
 \left.\frac{\partial n_0}{\partial e\bar
 V_\mathrm{g}}\right|_{F_\mathrm{e}=0}X,  
\end{equation}
i.e., the current-induced force generates 
a term in the effective potential which is quadratic in $X$.  
Far from the current-induced instability, this term gives a small renormalization
of the resonance frequency, 
$\Delta\omega_0/\omega_0
=(E^0_\mathrm{E}/2)
\left.{\partial n_0/\partial e\bar
V_\mathrm{g}}\right|_{F_\mathrm{e}=0}$,
from which we can extract a reliable estimate of 
the energy scale of the current blockade,
\begin{equation}
  E^0_\mathrm{E}=\frac{2C_\mathrm{g}}{C_\Sigma}\frac{\Delta\omega_0}{\omega_0}
  \left(\left.\frac{\partial n_0}{\partial e
  V_\mathrm{g}}\right|_{F_\mathrm{e}=0}\right)^{-1}.
\end{equation}
From the experiments of
Refs.~\onlinecite{steel09_Science} and \onlinecite{lassa09_Science}, we extract 
a value of ${\Delta\omega_0}/{\omega_0}$ of a few percents for
$V_\mathrm{g}$ at the degeneracy
point. 
The derivative of $n_0$ with respect to $e\bar V_\mathrm{g}$ can be estimated as the inverse of the 
width of the conductance peak in the $V$-$V_\mathrm{g}$ plane, divided by
$C_\mathrm{g}/C_\Sigma$.
This last quantity can, in turn, be estimated from the slope of the Coulomb diamonds.
Collecting these ingredients we find that for the suspended carbon
nanotubes of Ref.~\onlinecite{steel09_Science},  
$E^0_\mathrm{E}\simeq 3$--$5 \;\mu\mathrm{eV}$ which corresponds to 
$\tilde\alpha\simeq10^{-10}$, while for those or
Ref.~\onlinecite{lassa09_Science}, we get
$E^0_\mathrm{E}\simeq 20 \;\mu\mathrm{eV}$ and $\tilde\alpha\simeq10^{-8}$ (see
Ref.~\onlinecite{footnote:CNT}).

We now use these numbers to estimate the possible enhancement of the current
blockade near the Euler buckling instability. Based on Eq.~\eqref{eq:gap}, 
these parameters yield a possible increase of the mechanically-induced gap
by three orders of magnitude, leading to a maximal 
$\Delta_v$ (converted into a dimensionful quantity using the energy scale $E_\mathrm{E}^0$) 
of the order of 3 to \unit[5]{meV}.
Such large gaps would be much more easily observable in experiment. 
The implementation of such a device could be performed by 
the method routinely employed to control break junctions through a 
force pushing the substrate of the device.

We also
emphasize here again that it is preferable to operate the system near zero
excess charge on the quantum dot, 
where there are only a few electrons on the nanotube such that the
increase of the energy gap close to the Euler instability is not smeared out by
the additional force exerted on the nanotube (see
Sec.~\ref{sec:average_charge}). 
However, for the parameters of Refs.~\onlinecite{steel09_Science} and 
\onlinecite{lassa09_Science}, we estimate that the enhancement of the
current blockade remains very substantial for any realistic value of the excess
charge. 

When the tunneling-induced width $\Gamma$ becomes larger than or of the order of 
temperature, co-tunneling effects tend to smear the current blockade, \cite{pisto08_PRB,
koch06_PRB} as direct electronic transitions between left and right leads take
place.
However, close to the buckling instability the gap may remain larger than
temperature so that 
co-tunneling corrections should be suppressed in the immediate vicinity of the
instability. 

A last comment is in order on the required precision in the control of the
lateral compression force $F$.
As mentioned above, 
the increase of $\Delta_v$ is larger for smaller $\tilde\alpha$. But at the same
time, the increase is limited to a small force range.
The increase of the gap near the transition goes as $1/|F-F_\mathrm{c}|$. Thus 
when $\tilde\alpha$ is very small, a stringent requirement will be the precision in 
$F$ that we denote by $\delta F$. 
In this case the maximal gap will be of the order of 
$\frac{3}{4\pi}\frac{F_\mathrm{c}}{\delta F}\ln{\left(\frac{\delta
F/F_\mathrm{c}}{\sqrt[3]{\tilde \alpha}}\right)}$, assuming that $\delta
F/F_\mathrm{c}\gg\sqrt[3]{\tilde \alpha}$. 
This result can easily be checked by convolution of the gap \eqref{eq:gap} with a
Lorentzian of width $\delta F$.
This implies that if one is able to control the force with a precision sufficient 
to see the buckling instability ($\delta F/F_\mathrm{c} \ll 1$), 
there remains a large enhancement of the gap.

\section{Conclusions}
\label{sect:conclusions}
In this work, we have investigated the consequences of a capacitive
electromechanical coupling in a suspended single-electron transistor when the
supporting beam is brought close to the Euler buckling instability by a lateral
compression force. Our main result is that the low-bias current blockade
originating from the coupling between the electronic degrees of freedom and the
classical resonator can be enhanced by several orders of magnitude in the
vicinity of the instability. 
We show that both the mechanical as well as the electronic properties of this
regime can be described in an asymptotically exact manner based on a Langevin
equation.
These results are a direct consequence of the continuous nature of the Euler
buckling instability and the associated
``critical slowing down" of the fundamental bending mode of the beam at the
instability. 
In fact, more generally our results frequently have close and instructive
analogies with the mean-field theory of second-order phase transitions.
We focused on the sequential-tunneling transport
regime of single-level quantum dots, 
but many of our qualitative results should remain valid also
in the metallic case as well as for the resonant transport regime.
\cite{weick_unpublished}
In fact, our basic approach should apply quite generally for any
\textit{continuous} mechanical instability of a nano-electromechanical system.

Our result apply most directly to quantum dots situated on
nanobeams or carbon nanotubes. Applying strain to the nanobeam in a controlled
manner could, in principle, be experimentally performed with the help of a 
break junction.
In fact, it is quite conceivable that, e.g., some
carbon nanotube structures happen to be close to the Euler instability due to
specifics in the fabrication of individual nanostructures. Our predictions may
be helpful to identify such ``anomalous" (and potentially interesting) samples.

\begin{acknowledgments}
We acknowledge stimulating discussions with Eros Mariani, as
well as financial support by ANR contract JCJC-036 NEMESIS, and by the DFG
through Sfb 658. 
\end{acknowledgments}

\appendix
\section{Elasticity theory of the Euler instability}
\label{sec:Euler}
\begin{figure}[tb]
\includegraphics[width=.8\columnwidth]{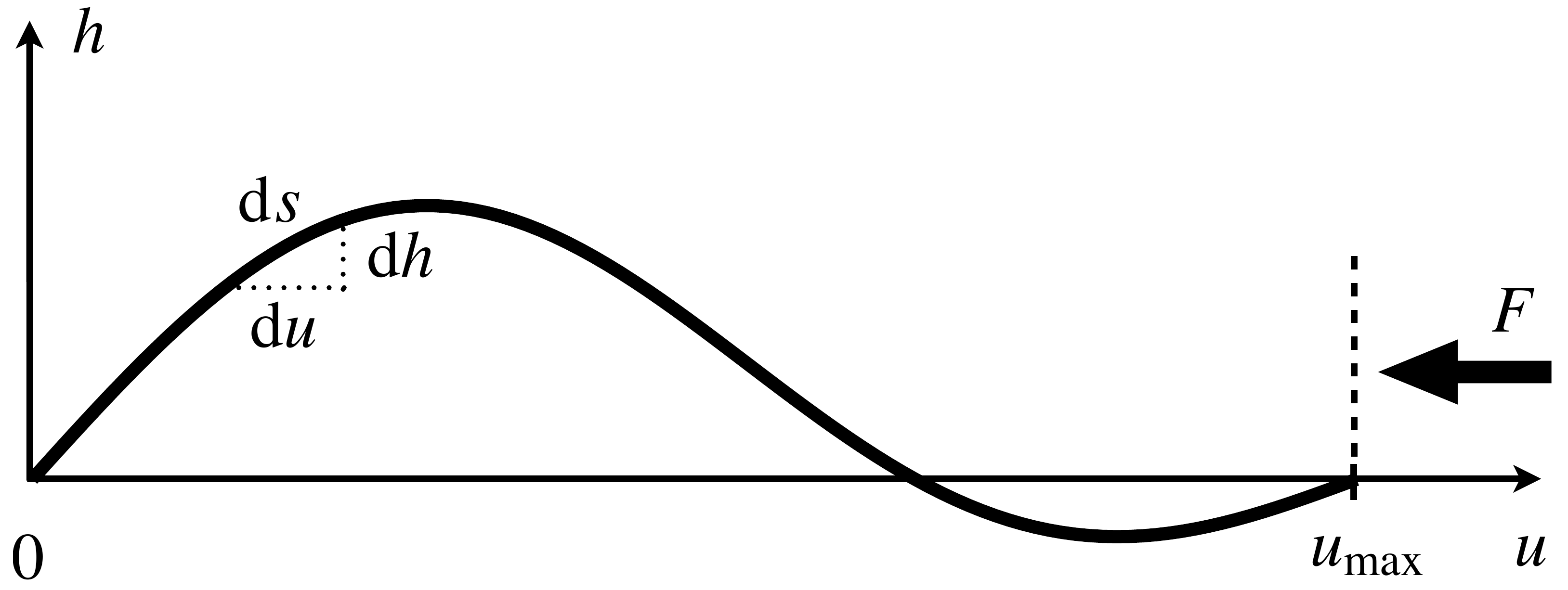}
\caption{\label{fig:rod}%
Coordinate system used to describe the elastic properties of the rod.}
\end{figure}

The elastic Lagrangian of a homogeneous rod of constant length $L$ 
fixed at its two ends consists of three parts, 
\begin{equation}
\label{eq:L}
\mathcal{L}=\mathcal{T}-\mathcal{V}_\mathrm{b}-\mathcal{V}_F.
\end{equation}
The kinetic term reads
\begin{equation}
\mathcal{T}=\frac{\sigma}{2}\int_0^L\mathrm{d}s\,\dot h^2, 
\end{equation}
where $\sigma$ is the linear mass density, $s$ the arc length along
the rod, and $h(s, t)$ the displacement of the rod with respect to the $u$-axis (see
Fig.~\ref{fig:rod}). 
The bending energy, controlled by the bending rigidity $\kappa$, is given by
\begin{equation}
\label{eq:V_b}
\mathcal{V}_\mathrm{b}=\frac{\kappa}{2}\int_0^L\mathrm{d}s
\left|\frac{\mathrm{d}\hat t}{\mathrm{d}s}\right|^2
=\frac{\kappa}{2}\int_0^L\mathrm{d}s
\frac{{h''}^2}{1-{h'}^2},
\end{equation}
where $\hat t=(u', h')$ is the tangent vector of the rod and primes denote
derivative with respect to $s$.
The last term in Eq.~\eqref{eq:L} corresponds to the work done by the
compression force $F$ on the rod and reads
\begin{equation}
\mathcal{V}_F=-F(L-u_\mathrm{max})
=-F\int_0^L\mathrm{d}s\left(1-\sqrt{1-{h'}^2}\right),
\end{equation}
where $u_\mathrm{max}$ is the total extent of the rod along the $u$-axis
(see Fig.~\ref{fig:rod}).

For small deflections ($|h'|\ll1$), the Lagrangian \eqref{eq:L} becomes, in
harmonic approximation, 
\begin{equation}
\mathcal{L}\simeq\int_0^L\mathrm{d}s
\left(
\frac{\sigma}{2}{\dot h}^2-\frac{\kappa}{2}{h''}^2+\frac{F}{2}{h'}^2
\right)
\end{equation}
with the corresponding Euler-Lagrange equation 
\begin{equation}
\label{eq:Euler-Lagrange}
\sigma\ddot h+\kappa h''''+Fh''=0.
\end{equation}
Equation \eqref{eq:Euler-Lagrange} can be solved by the eigenfunctions $h(s,
t)=\sum_nh_n(s, t)=\sum_nX_n(t)g_n(s)$, where $g_n(s)$ are the normal modes
which follow from the solution of the characteristic equation. 
The frequency of the mode $n$ reads
\begin{equation}
\omega_n^2=\frac{\kappa}{\sigma}q_n^2\left(q_n^2-\frac{F}{\kappa}\right)
\end{equation}
with $q_n$ the associated wavenumber which depends on the
considered
boundary conditions. The
vibrational frequency of the fundamental bending mode ($n=1$) thus vanishes at the critical
force $F_\mathrm{c}=\kappa q_1^2$, while all higher modes have a finite
frequency and hence are neglected in what follows.
For $F>F_\mathrm{c}$, the fundamental mode is
unstable, and quartic corrections to the Lagrangian are necessary to ensure
global stability. Denoting $\omega_1=\omega$ and $X_1=X$,
expanding the Lagrangian \eqref{eq:L} to quartic order in the displacement and
inserting the solution $h_1$ of the harmonic problem, we
thus obtain the effective Lagrangian close to the Euler instability, 
\begin{equation}
\label{eq:L_eff}
\mathcal{L}=\frac{m}{2}\dot X^2-\frac{m\omega^2}{2}X^2-\frac{\alpha}{4}X^4, 
\end{equation}
with the effective mass
\begin{equation}
m=\sigma\int_0^L\mathrm{d}s\,g_1^2,
\end{equation}
and 
\begin{equation}
\label{eq:alpha}
\alpha=\int_0^L\mathrm{d}s\left(2\kappa
{g_1''}^2{g_1'}^2-\frac{F_\mathrm{c}}{2}{g_1'}^4\right),
\end{equation}
with $g_1(L/2)=1$ such that $X$ corresponds in Eq.~\eqref{eq:L_eff} to the actual displacement of the center
of the rod.
Notice that a priori, $\alpha$ depends on the force $F$. However, close to the
buckling instability, we can approximate $F\simeq F_\mathrm{c}$ in
Eq.~\eqref{eq:alpha}.

The parameters entering the effective Lagrangian \eqref{eq:L_eff} and the corresponding
vibrational Hamiltonian \eqref{eq:H_vib} are given in Table
\ref{tab:parameters} for two types of boundary conditions: 
hinged end points 
($h\big|_{0,L}=h''\big|_{0,L}=0$) and clamped end points
($h\big|_{0,L}=h'\big|_{0,L}=0$). 
\begin{table*}[bth] 
\caption{\label{tab:parameters}Parameters entering the effective Lagrangian
\eqref{eq:L_eff} and the vibrational Hamiltonian \eqref{eq:H_vib}.}
\begin{ruledtabular} 
\begin{tabular}{cccccc}
Boundary conditions & $g_1(s)$ & $F_\mathrm{c}$ & $\omega^2$ & $m$ & $\alpha$
\vspace{.2cm}\\
$h\big|_{0,L}=h''\big|_{0,L}=0$ & $\displaystyle\sin{\left(\frac{\pi s}{L}\right)}$ 
& $\displaystyle\kappa\left(\frac{\pi}{L}\right)^2$ 
& $\displaystyle\frac{\kappa}{\sigma}\left(\frac{\pi}{L}\right)^4\left(1-\frac{F}{F_\mathrm{c}}\right)$ 
& $\displaystyle\frac{\sigma L}{2}$ & $\displaystyle F_\mathrm{c}L\left(\frac{\pi}{2L}\right)^4$
\vspace{.2cm}\\
$h\big|_{0,L}=h'\big|_{0,L}=0$ & $\displaystyle\approx\sin^2{\left(\frac{\pi s}{L}\right)}$ 
& $\displaystyle\kappa\left(\frac{2\pi}{L}\right)^2$ 
& $\displaystyle\frac{\kappa}{\sigma}\left(\frac{2\pi}{L}\right)^4\left(1-\frac{F}{F_\mathrm{c}}\right)$ 
& $\displaystyle\frac{3\sigma L}{8}$ & $\displaystyle F_\mathrm{c}L\left(\frac{\pi}{2L}\right)^4$ 
\end{tabular} 
\end{ruledtabular} 
\end{table*}
Notice that in the latter case, only an
approximate solution of the Euler-Lagrange equation \eqref{eq:Euler-Lagrange}
can be found, which is valid in the vicinity of the Euler
instability, i.e., for $F\simeq F_\mathrm{c}$.

\section{Langevin dynamics of the mechanical degree of freedom}
\label{sec:FP}
For the convenience of the reader, we present a derivation of the Fokker-Planck
equation \eqref{eq:FP}. Our derivation is quite general as long as electronic transport is
described by rate equations (sequential tunneling). We note that Fokker-Planck
equations for nano-electromechanical systems appeared previously, e.g., in
Ref.~\onlinecite{blant04_PRL}. 

We assume that the electromechanical coupling takes the general form
$H_\mathrm{c}=h(X)n_\mathrm{d}$, 
where $h$ is an arbitrary function of the mode amplitude $X$. 
In the classical limit ($\hbar|\omega|\ll k_\mathrm{B}T$), and in the sequential
tunneling regime ($\hbar\Gamma\ll k_\mathrm{B}T$), 
one can write a Boltzmann equation for the joint probability distribution 
$\mathcal{P}_n(X, P, t)$ that the resonator is in charge state $n$ ($=0,1$) and
phase space point $(X, P)$ at time $t$ (Refs.~\onlinecite{armou04_PRB}, \onlinecite{elste08_APA}),
\begin{equation}
\label{eq:Boltzmann_equation}
\partial_t\mathcal{P}_n=\left\{\mathcal{H}_n, \mathcal{P}_n\right\}
-(-1)^n\Gamma_{01}(X)\mathcal{P}_0+(-1)^n\Gamma_{10}(X)\mathcal{P}_1.
\end{equation}
The Poisson bracket $\{f, g\}=\partial_X f\partial_P g-\partial_P f\partial_X
g$ describes the classical dynamics of the mechanical degree of freedom on the adiabatic
potentials corresponding to the neutral and singly-charged states of the quantum
dot with Hamiltonian $\mathcal{H}_n=H_\mathrm{vib}+h(X)n$,
where $H_\mathrm{vib}$ is the vibrational Hamiltonian of Eq.~\eqref{eq:H_vib}.
In Eq.~\eqref{eq:Boltzmann_equation}, the position-dependent rates for tunneling
of electrons in and 
out of the dot, $\Gamma_{01}(X)$ and $\Gamma_{10}(X)$, respectively, account for the
electronic dynamics. Notice that the following derivation does not depend on the specific
form of these rates, and hence on the particular model for the quantum dot which one considers (e.g.,
metallic or molecular).

Close to the Euler instability, the vibrational mode becomes slow (``critical
slowing down") and the 
Poisson bracket in Eq.~\eqref{eq:Boltzmann_equation} can be considered as a
small perturbation. 
If we neglect the Poisson bracket entirely in a first step, the stationary solution of
Eq.~\eqref{eq:Boltzmann_equation} reads
$\mathcal{P}_0=\Gamma_{10}\mathcal{P}/\Gamma$,
$\mathcal{P}_1=\Gamma_{01}\mathcal{P}/\Gamma$, with
$\mathcal{P}=\mathcal{P}_0+\mathcal{P}_1$ and $\Gamma=\Gamma_{01}+\Gamma_{10}$.
Next, we account for the Poisson bracket perturbatively to leading order by
making the ansatz
\begin{subequations}
\label{eq:ansatz}
\begin{align}
\mathcal{P}_0(X, P, t)&=\frac{\Gamma_{10}(X)}{\Gamma(X)}\mathcal{P}(X, P, t)
-\delta\mathcal{P}(X, P, t),\\
\mathcal{P}_1(X, P, t)&=\frac{\Gamma_{01}(X)}{\Gamma(X)}\mathcal{P}(X, P, t)
+\delta\mathcal{P}(X, P, t).
\end{align}
\end{subequations}
In the adiabatic limit, in which electronic tunneling is much faster than the
vibrational dynamics ($|\omega|\ll\Gamma$), one would then expect that $\delta\mathcal{P}$ is small compared
to $\mathcal{P}$ itself. 
Inserting Eq.~\eqref{eq:ansatz} into Eq.~\eqref{eq:Boltzmann_equation}, we obtain
\begin{equation}
\label{eq:P}
\partial_t\mathcal{P}=-\frac{P}{m}\partial_X\mathcal{P}-F_\mathrm{eff}(X)\partial_P\mathcal{P}
+\left(\partial_Xh\right)\partial_P\delta\mathcal{P}
\end{equation}
with the effective force $F_\mathrm{eff}(X)=-\partial_XH_\mathrm{vib}
-\left(\partial_Xh\right)n_0(X)$ and $n_0(X)=\Gamma_{01}(X)/\Gamma(X)$ the average
occupation of the dot for fixed position $X$, and 
\begin{align}
\label{eq:delta_P}
\partial_t\delta\mathcal{P}=&\,\left\{\mathcal{H}_0,
\delta\mathcal{P}\right\}-\Gamma\delta\mathcal{P}
+\left(\partial_Xh\right)\left(\frac{\Gamma_{01}\Gamma_{10}}{\Gamma^2}\partial_P\mathcal{P}
+\frac{\Gamma_{10}}{\Gamma}\partial_P\delta\mathcal{P}\right)
\nonumber\\
&-\frac{P}{m}\frac{\Gamma_{10}\partial_X\Gamma_{01}-\Gamma_{01}\partial_X\Gamma_{10}}{\Gamma^2}\mathcal{P}.
\end{align}

So far, Eqs.~\eqref{eq:P} and \eqref{eq:delta_P} are exact, and we now make
use of the adiabatic limit $\delta\mathcal{P}\ll\mathcal{P}$ to solve them. In
this limit, all terms in Eq.~\eqref{eq:delta_P} containing $\delta\mathcal{P}$
are negligible compared to those involving $\mathcal{P}$, except for the
second term on the right-hand side which is multiplied by the total tunneling
rate $\Gamma\gg|\omega|$. We thus obtain from Eq.~\eqref{eq:delta_P}
\begin{equation}
\delta\mathcal{P}\simeq
\left(\partial_Xh\right)\frac{\Gamma_{01}\Gamma_{10}}{\Gamma^3}\partial_P\mathcal{P}
-\frac{P}{m}\frac{\Gamma_{10}\partial_X\Gamma_{01}-\Gamma_{01}\partial_X\Gamma_{10}}{\Gamma^3}\mathcal{P}.
\end{equation}
Inserting this expression into Eq.~\eqref{eq:P}, we obtain the Fokker-Planck
equation \eqref{eq:FP} (except for the purely extrinsic dissipative and
diffusive parts proportional to the damping constant $\eta_\mathrm{e}$, 
which can readily be obtained by coupling the mechanical degree
of freedom to a phonon bath), 
with damping coefficient
$\eta(X)=-\left(\partial_Xh\right)\left(\partial_Xn_0\right)/\Gamma$, 
and diffusion constant
$D(X)=2\left(\partial_Xh\right)^2n_0(1-n_0)/\Gamma$.

\section{Current blockade for a harmonic oscillator linearly coupled to a
SET}
\label{sec:harmonic}
In this Appendix, we detail the derivation of the Coulomb diamond and the
resulting current blockade at low bias voltage within the zero-temperature mean-field
approximation of Sec.~\ref{sect:mean-field}, in the specific case where the
resonator coupled to
the SET is purely harmonic. We follow and adapt Ref.~\onlinecite{pisto07_PRB} to
the case of a single-level quantum dot, where the average occupation of the
island is given by Eq.~\eqref{eq:n0}. Specifically, we write the effective potential as
\begin{equation}
v_\mathrm{eff}(x)=v(x)+\int^x{\rm d}x'n_0(x'), 
\end{equation}
with
\begin{equation}
v(x)=\frac{v''(\bar x)}{2}\left(x-\bar x\right)^2. 
\end{equation}
Introducing $\tilde x=x-\bar x$ and $\tilde v_\mathrm{g}=v_\mathrm{g}-\bar x$, and using
Eq.~\eqref{eq:n0}, the condition \eqref{eq:equilibrium}
for dynamical equilibrium at zero temperature reads
\begin{equation}
\label{eq:dynamical_eq}
-v''(\bar x)\tilde x=
\begin{cases}
1, & \displaystyle\tilde x<\tilde v_\mathrm{g}-\frac v2,\vspace{.2truecm}\\
\displaystyle\frac 12, & \displaystyle\tilde v_\mathrm{g}-\frac v2\leqslant\tilde x\leqslant\tilde
v_\mathrm{g}+\frac v2,\vspace{.2truecm}\\
0, & \displaystyle\tilde x>v_\mathrm{g}+\frac v2.
\end{cases}
\end{equation}

\begin{figure}[tb]
\includegraphics[width=.75\columnwidth]{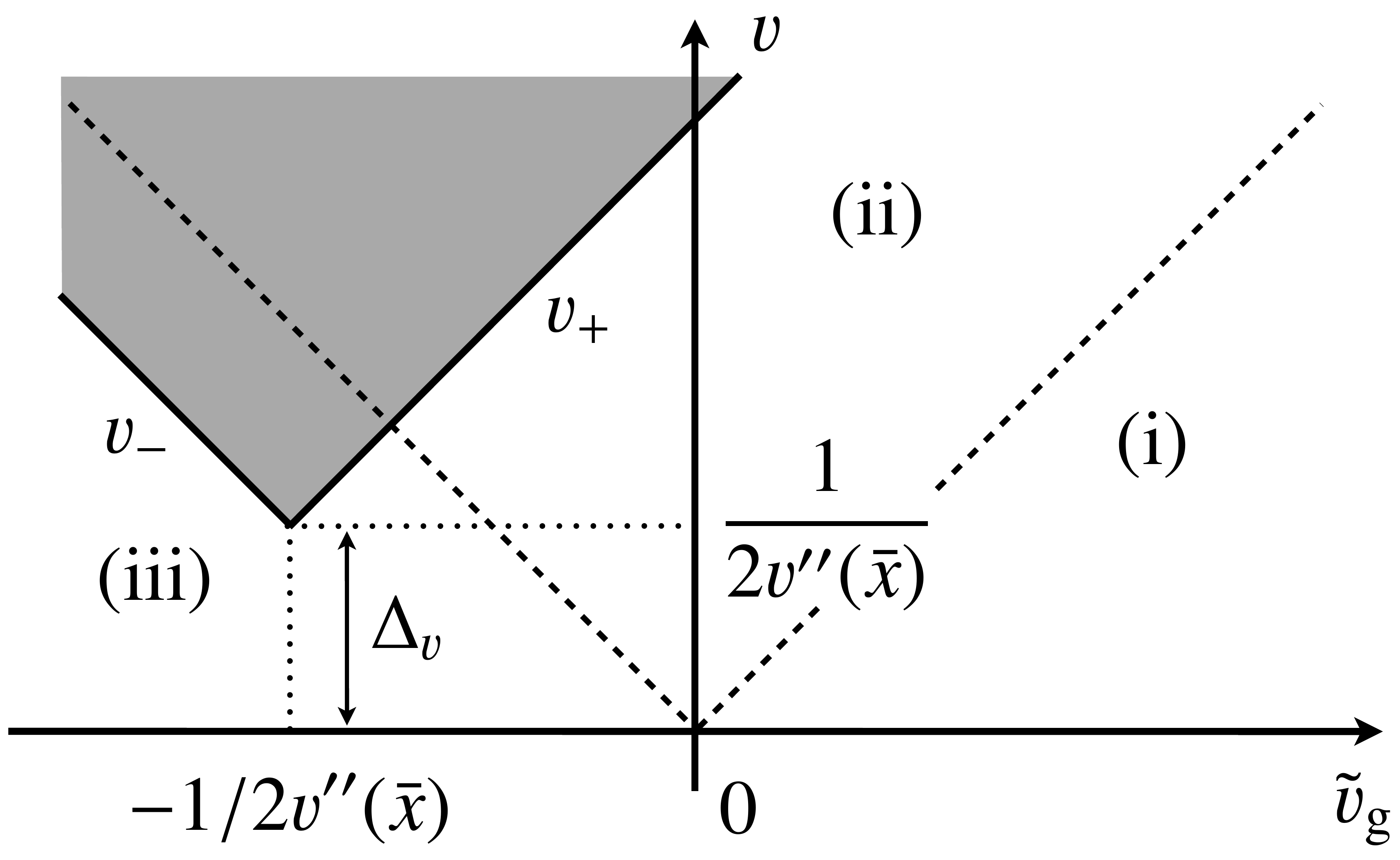}
\caption{\label{fig:sketch_CB_diamond}%
Sketch of the Coulomb diamond for a SET linearly coupled to a harmonic
oscillator, delimited by the solid lines at $v_+=2\tilde
v_\mathrm{g}+3/2v''(\bar x)$ and $v_-=-2\tilde v_\mathrm{g}-1/2v''(\bar x)$.
The gray area indicates the region where current can flow.
The dashed lines indicate the location of the 
Coulomb diamond without electromechanical coupling ($F_\mathrm{e}=0$), delimited by $v=\pm
2\tilde v_\mathrm{g}$, which defines the three regions of the $v$-$\tilde
v_\mathrm{g}$ plane discussed in the text.}
\end{figure}
In order to solve Eq.~\eqref{eq:dynamical_eq}, one has to distinguish three
cases (see Fig.~\ref{fig:sketch_CB_diamond}):
\begin{itemize}
\item[(i)] $\tilde v_\mathrm{g}>v/2$: In that case, there only exists one
solution to Eq.~\eqref{eq:dynamical_eq} which corresponds to $n_0=1$ and thus is
not conducting, $\tilde x_1=-1/v''(\bar x)$. Therefore, for $\tilde
v_\mathrm{g}>v/2$, the system is always non conducting.
\item[(ii)] $|\tilde v_\mathrm{g}|<v/2$: Here, two solutions can coexist: $\tilde x_1$ (non
conducting), and $\tilde x_{1/2}=-1/2v''(\bar x)$ which corresponds to $n_0=1/2$
and hence represents the conducting state of the effective potential. In order
for the system to be conducting within our mean-field approximation, this latter
solution must be the most stable. This happens whenever $v>
v_+\equiv2\tilde v_\mathrm{g}+3/2v''(\bar x)$.
\item[(iii)] $\tilde v_\mathrm{g}<-v/2$: In this last case, an additional solution to
Eq.~\eqref{eq:dynamical_eq} exists on top of $\tilde x_1$ and $\tilde x_{1/2}$,
namely $\tilde x_0=0$ which is not conducting and corresponds to $n_0=0$. The
conducting solution $\tilde x_{1/2}$ is the most stable one if $v>v_+$ and
$v>v_-\equiv-2\tilde v_\mathrm{g}-1/2v''(\bar x)$.
\end{itemize}
The apex (defined by $v_+=v_-$) of the resulting Coulomb diamond sketched in
Fig.~\ref{fig:sketch_CB_diamond} is thus located
at a bias voltage 
\begin{equation}
\Delta_v=\frac{1}{2v''(\bar x)}
\end{equation}
which defines the energy gap below
which current cannot flow through the system. The gate voltage corresponding to
such a gap is given by $v_\mathrm{g}=-1/2v''(\bar x)+\bar x$.

\section{Analytical treatment of thermal fluctuations}
\label{app:temp}
We present here a detailed analytical treatment of the 
behavior of the current when one only considers thermal fluctuations in the
Fokker-Planck equation \eqref{eq:scaled_FP}, and compare it to 
the numerical results presented in Fig.~\ref{fig:extrinsic_dissipation} of Sec.~\ref{sec:temp}.

For temperatures much smaller than the gap, the onset of the current with bias
voltage has a Fermi-function-like behavior (see dashed and
dashed-dotted lines in Fig.~\ref{fig:extrinsic_dissipation}). This behavior can readily be
checked by approximating the effective potential \eqref{eq:v_eff} for gate
voltages at the apex of the Coulomb diamond [see Eq.~\eqref{eq:vg}] by its
zero-temperature expression, and by expanding the Boltzmann distribution
\eqref{eq:Boltzmann} close to the corresponding minima of that potential. In
this limit, the Boltzmann distribution is a superposition of weighted Gaussian
peaks centered at these minima (except for $|\delta|\ll\sqrt[3]{\alpha}$ and $x\simeq0$, where the
effective potential is purely quartic in $x$). For
$\tilde T\ll\Delta_v$, 
we find from Eq.~\eqref{eq:I} for the current
\begin{equation}
\label{eq:I_lowT}
I\simeq\frac{e\Gamma}{4}\left[g\left(\frac{v}{\Delta_v}\right)
\exp{\left(\frac{\Delta_v-v}{4\tilde T}\right)}+1\right]^{-1},
\end{equation}
in good agreement with our numerical results presented in Sec.~\ref{sec:temp}.
Far from the Euler instability
($|\delta|\gg\sqrt[3]{\alpha}$), the function $g(z)$ is given by
\begin{equation}
\label{eq:g1}
g(z)=
\begin{cases}
2, & 0<z<2,\\
0, & z\geqslant2.
\end{cases}
\end{equation}
In the vicinity of the instability ($|\delta|\ll\sqrt[3]{\alpha}$), we have
\begin{equation}
\label{eq:g2}
g(z)=
\begin{cases}
\infty, & 
0<z<z_1,
\vspace{.2truecm}\\
\displaystyle
2^{-1/3}+\frac{2^{2/3}\Gamma(5/2)}{\sqrt{\pi}}\left(\frac{6\Delta_v}{\tilde T}\right)^{1/4}, & 
z_1\leqslant z<z_2,
\vspace{.2truecm}\\
\displaystyle
\frac{2^{2/3}\Gamma(5/2)}{\sqrt{\pi}}\left(\frac{6\Delta_v}{\tilde T}\right)^{1/4}, & 
z_2\leqslant z<z_3,
\vspace{.2truecm}\\
0, &z\geqslant z_3, 
\end{cases}
\end{equation}
where $\Gamma(\nu)$ denotes the gamma function, and 
$z_1=(2^{5/3}/3-1)/(1-2^{-1/3})$, $z_2=1/3(1-2^{-1/3})$, and
$z_3=1/(1-2^{-1/3})$.
Notice that the discontinuities in Eqs.~\eqref{eq:g1} and \eqref{eq:g2} are
due to the fact that (meta)stable conducting or blocked minima of the effective potential are
appearing or disappearing as one increases the bias voltage. 
Our approximate result \eqref{eq:I_lowT} shows that at low temperature,
the current below the gap is exponentially suppressed as a function of bias
voltage. 

At larger temperatures, the current starts to be linear in the bias voltage, as can be seen
from the dotted line
in Fig.~\ref{fig:extrinsic_dissipation} for $\tilde T/\Delta_v=1$. 
Indeed, expanding for $\tilde T\gg|v_\mathrm{g}\pm v/2|$ 
the current for fixed $x$, 
Eq.~\eqref{eq:I(X)}, as well as the Boltzmann distribution \eqref{eq:Boltzmann}, we
find using Eq.~\eqref{eq:I}
\begin{equation}
\label{eq:I_highT}
I\simeq C\frac{e\Gamma}{4}\frac{v}{\tilde T}, 
\end{equation}
with 
\begin{equation}
C=\frac{\displaystyle\int \mathrm{d}y f_\mathrm{F}(y)
\exp{\left(\frac{\delta \tilde T}{2}y^2-\frac{\tilde\alpha \tilde T^3}{4}y^4\right)}}
{\displaystyle\int \mathrm{d}y f_\mathrm{F}^{-1}(-y)
\exp{\left(\frac{\delta \tilde T}{2}y^2-\frac{\tilde\alpha \tilde
T^3}{4}y^4\right)}},
\end{equation}
to first non-vanishing order in $(v, |v_\mathrm{g}|)/\tilde T$. For $\tilde
T\gg\Delta_v$, we find $C\simeq1/4$, such
that 
\begin{equation}
I\simeq e\Gamma \frac{eV}{16 k_\mathrm{B} T},
\end{equation}
which corresponds to the usual high-temperature current in absence of
electromechanical coupling.
We have checked that this result is in very
good agreement with our numerical calculation of Sec.~\ref{sec:temp}. 
Of course,
when the bias voltage becomes significantly larger than temperature, the current saturates to its
maximal value $I=e\Gamma/4$.

\section{Stationary solution of the Fokker-Planck equation}
\label{sec:scaling}
We show here that in the adiabatic limit ($\omega_0/\Gamma\ll1$)
and for weak extrinsic dissipation ($\gamma_\mathrm{e}\ll1$), the stationary solution of the Fokker-Planck
equation \eqref{eq:scaled_FP} only depends on the ratio
$\frac{\gamma_\mathrm{e}}{\omega_0/\Gamma}$. This behavior is examplified in
Fig.~\ref{fig:fluctuations} where the averaged current flowing through the
nanobeam only depends on the latter ratio.

Introducing the new variables
\begin{align}
E(x, p)&=\frac{p^2}{2}+v_\mathrm{eff}(x),\\
\theta(x, p)&=\int_{x_0}^x\frac{\mathrm{d}x'}{\dot x'}
=\int_{x_0}^x\frac{\mathrm{d}x'}{\sqrt{2[E(x, p)-v_\mathrm{eff}(x')]}},
\end{align}
with $E$ the energy of the mechanical degree of freedom and $\theta$ the time along a trajectory in phase
space at a given $E$ ($x_0$ is the initial position of the system for that
particular trajectory), the Fokker-Planck equation
\eqref{eq:scaled_FP} reads
\begin{align}
\label{eq:FP_transformed}
\frac{\partial\mathcal{P}}{\partial\tau}=&-\frac{\partial\mathcal{P}}{\partial\theta}
+\left(\gamma(x)+\gamma_\mathrm{e}\right)
\left(\mathcal{P}+p^2\frac{\partial\mathcal{P}}{\partial E}
+p\frac{\partial\theta}{\partial p}\frac{\partial\mathcal{P}}{\partial\theta}\right)
\nonumber\\
&+\left(\frac{d(x)}{2}+\gamma_\mathrm{e}\tilde T\right)
\Bigg(\frac{\partial\mathcal{P}}{\partial E}+p^2\frac{\partial^2\mathcal{P}}{\partial E^2}
+\frac{\partial^2\theta}{\partial p^2}\frac{\partial\mathcal{P}}{\partial\theta}
\nonumber\\
&+\left(\frac{\partial\theta}{\partial p}\right)^2\frac{\partial^2\mathcal{P}}{\partial\theta^2}
+2p\frac{\partial\theta}{\partial p}\frac{\partial^2\mathcal{P}}{\partial
E\partial\theta}\Bigg).
\end{align}
Noticing that the current-induced fluctuation and dissipation are both
proportional to $\omega_0/\Gamma$ [cf.\ Eqs.~\eqref{eq:d} and \eqref{eq:gamma}],
we obtain that for $\omega_0/\Gamma=\gamma_\mathrm{e}=0$, the stationary solution
of Eq.~\eqref{eq:FP_transformed}, $\partial_\tau\mathcal{P}_\mathrm{st}=0$, is independent of $\theta$. This suggests to
insert the ansatz
\begin{equation}
\mathcal{P}_\mathrm{st}(E,
\theta)=\bar{\mathcal{P}}_\mathrm{st}(E)+\delta\mathcal{P}_\mathrm{st}(E, \theta)
\end{equation}
with $\delta\mathcal{P}_\mathrm{st}\ll\bar{\mathcal{P}}_\mathrm{st}$ in
Eq.~\eqref{eq:FP_transformed}. To first order in $(\omega_0/\Gamma,
\gamma_\mathrm{e})\ll1$, we obtain
\begin{align}
0=&-\frac{\partial\delta\mathcal{P}_\mathrm{st}}{\partial\theta}
+\left(\gamma(x)+\gamma_\mathrm{e}\right)
\left(\bar{\mathcal{P}}_\mathrm{st}+p^2\frac{\partial\bar{\mathcal{P}}_\mathrm{st}}{\partial E}
\right)
\nonumber\\
&+\left(\frac{d(x)}{2}+\gamma_\mathrm{e}\tilde T\right)
\left(\frac{\partial\bar{\mathcal{P}}_\mathrm{st}}{\partial E}
+p^2\frac{\partial^2\bar{\mathcal{P}}_\mathrm{st}}{\partial E^2}
\right).
\end{align}
Averaging this equation over one period $\mathcal{T}$ of the motion in phase
space, and using the periodicity of $\delta\mathcal{P}_\mathrm{st}$ in $\theta$, we have
\begin{align}
\label{eq:FP_E}
0=&\left<\left(\frac{\gamma(x)}{\gamma_\mathrm{e}}+1\right)\left(1+p^2\frac{\mathrm{d}}{\mathrm{d}E}\right)\right>_E
\bar{\mathcal{P}}_\mathrm{st}
\nonumber\\
&+\left<\left(\frac{d(x)}{2\gamma_\mathrm{e}}+\tilde T\right)\left(1+p^2\frac{\mathrm{d}}{\mathrm{d}E}\right)\right>_E
\frac{\mathrm{d}\bar{\mathcal{P}}_\mathrm{st}}{\mathrm{d}E},
\end{align}
where
\begin{equation}
\left<f(x, p)\right>_E=\frac{1}{\mathcal{T}}\int_0^\mathcal{T}\mathrm{d}\theta
f(x, p).
\end{equation}
Since $\gamma(x)$ and $d(x)$ only depends on $\omega_0/\Gamma$ via their
prefactors, it is clear from Eq.~\eqref{eq:FP_E} that the stationary solution of
the Fokker-Planck equation only depends on $\omega_0$, $\Gamma$, and
$\gamma_\mathrm{e}$ through the ratio
$\frac{\gamma_\mathrm{e}}{\omega_0/\Gamma}$.



\begin{thebibliography}{}

\bibitem{knobe02_APL}
R.\ G.\ Knobel and A.\ N.\ Cleland, 
Appl.\ Phys.\ Lett.\ \textbf{81}, 2258 (2002).

\bibitem{knobe03_Nature}
R.\ G.\ Knobel and A.\ N.\ Cleland, 
Nature (London) \textbf{424}, 291 (2003).

\bibitem{armou02_PRL}
A.\ D.\ Armour, M.\ P.\ Blencowe, and K.\ C.\ Schwab, 
Phys.\ Rev.\ Lett.\ \textbf{88}, 148301 (2002).

\bibitem{gorel98_PRL}
L.\ Y.\ Gorelik, A.\ Isacsson, M.\ V.\ Voinova, B.\ Kasemo, R.\ I.\ Shekhter,
and M.\ Jonson, 
Phys.\ Rev.\ Lett.\ \textbf{80}, 4526 (1998).

\bibitem{pisto05_PRL}
F.\ Pistolesi and R.\ Fazio, 
Phys.\ Rev.\ Lett.\ \textbf{94}, 036806 (2005).

\bibitem{usman07_PRB}
O.\ Usmani, Ya.\ M.\ Blanter, and Yu.\ V.\ Nazarov, 
Phys.\ Rev.\ B \textbf{75}, 195312 (2007). 

\bibitem{pisto07_PRB}
F.\ Pistolesi and S.\ Labarthe, 
Phys.\ Rev.\ B \textbf{76}, 165317 (2007). 

\bibitem{koch05_PRL}
J.\ Koch and F.\ von Oppen, 
Phys.\ Rev.\ Lett.\ \textbf{94}, 206804 (2005).

\bibitem{koch06_PRB}
J.\ Koch, F.\ von Oppen, and A.\ V.\ Andreev, 
Phys.\ Rev.\ B \textbf{74}, 205438 (2006).

\bibitem{letur09_NaturePhysics}
R.\ Leturcq, C.\ Stampfer, K.\ Inderbitzin, L.\ Durrer, C.\ Hierold, E.\
Mariani, M.\ G.\ Schultz, F.\ von Oppen, and K.\ Ensslin, 
Nat.\ Phys.\ \textbf{5}, 327 (2009).

\bibitem{mozyr06_PRB} 
D.\ Mozyrsky, M.\ B.\ Hastings, and I.\ Martin, 
Phys.\ Rev.\ B \textbf{73}, 035104 (2006).

\bibitem{pisto08_PRB}
F.\ Pistolesi, Ya.\ M.\ Blanter, and I.\ Martin, 
Phys.\ Rev.\ B \textbf{78}, 085127 (2008).

\bibitem{armou04_PRB}
A.\ D.\ Armour, M.\ P.\ Blencowe, and Y.\ Zhang, 
Phys.\ Rev.\ B \textbf{69}, 125313 (2004).

\bibitem{blant04_PRL} 
Ya.~M.\ Blanter, O.\ Usmani, and Yu.~V.\ Nazarov,
Phys.\ Rev.\ Lett.\ \textbf{93}, 136802 (2004);
\textbf{94}, 049904(E) (2005).

\bibitem{doiro06_PRB}
C.\ B.\ Doiron, W.\ Belzig, and C.\ Bruder, 
Phys.\ Rev.\ B \textbf{74}, 205336 (2006).

\bibitem{steel09_Science} 
G.\ A.\ Steele, A.\ K.\ H\"uttel, B.\ Witkamp, M.\ Poot, H.\ B.\ Meerwaldt, 
L.\ P.\ Kouwenhoven, and H.\ S.\ J.\ van der Zant, 
Science {\bf 325}, 1103 (2009).

\bibitem{lassa09_Science} 
B.\ Lassagne, Y.\ Tarakanov, J.\ Kinaret, D.\ Garcia-Sanchez, and A.\ Bachtold, 
Science {\bf 325}, 1107 (2009).

\bibitem{euler}
L.\ Euler, in \textit{Leonhard Euler's Elastic Curves}, translated and annotated
by W.~A.\ Oldfather, C.~A.\ Ellis, and D.~M.\ Brown, reprinted from ISIS, No.~58
XX(1), 1744 (Saint Catherine Press, Bruges).

\bibitem{landau}
L.\ D.\ Landau and E.\ M.\ Lifshitz, 
\textit{Theory of Elasticity} 
(Pergamon Press, Oxford, 1970).

\bibitem{falvo97_Nature}
M.\ R.\ Falvo, G.\ J.\ Harris, R.\ M.\ Taylor II, V.\ Chi, F.\ P.\ Brooks Jr,
S.\ Washburn, and R.\ Superfine, 
Nature (London) \textbf{389}, 582 (1997).

\bibitem{carr03_APL}
S.\ M.\ Carr and M.\ N.\ Wybourne, 
Appl.\ Phys.\ Lett.\ \textbf{82}, 709 (2003).

\bibitem{carr05_EPL}
S.\ M.\ Carr, W.\ E.\ Lawrence, and M.\ N.\ Wybourne, 
Europhys.\ Lett.\ \textbf{69}, 952 (2005).

\bibitem{roode09_APL}
D.\ Roodenburg, J.\ W.\ Spronck, H.\ S.\ J.\ van der Zant, and W.\ J.\ Venstra, 
Appl.\ Phys.\ Lett.\ \textbf{94}, 183501 (2009).

\bibitem{carr01_PRB}
S.\ M.\ Carr, W.\ E.\ Lawrence, and M.\ N.\ Wybourne, 
Phys.\ Rev.\ B \textbf{64}, 220101(R) (2001).

\bibitem{werne04_EPL}
P.\ Werner and W.\ Zwerger, 
Europhys.\ Lett.\ \textbf{65}, 158 (2004).

\bibitem{peano06_NJP}
V.\ Peano and M.\ Thorwart, 
New J.\ Phys.\ \textbf{8}, 21 (2006).

\bibitem{savel06_NJP}
S.\ Savel'ev, X.\ Hu, and F.\ Nori, 
New J.\ Phys.\ \textbf{8}, 105 (2006).

\bibitem{weick10_PRB}
G.\ Weick, F.\ Pistolesi, E.\ Mariani, and F.\ von Oppen, 
Phys.\ Rev.\ B \textbf{81}, 121409(R) (2010).

\bibitem{hutte09_NL}
A.\ K.\ H\"uttel, G.\ A.\ Steele, B.\ Witkamp, M.\ Poot, L.\ P.\ Kouwenhoven, 
and H.\ S.\ J.\ van der Zant,
Nano Lett.\ \textbf{9}, 2547 (2009).

\bibitem{footnote:CNT}
Typical parameters for carbon nanotubes of length $L$ and radius $R$ are:
$\kappa=6.3R^3[\mathrm{nm}]\,\mu\mathrm{eV.m}$,
$\sigma=4.2\times10^{-15}R[\mathrm{nm}]\,\mathrm{kg.m}^{-1}$, such that
$m=1.6\times10^{-21}R[\mathrm{nm}]L[\mu\mathrm{m}]\,\mathrm{kg}$, 
$\omega_0=610R[\mathrm{nm}]L^{-2}[\mu\mathrm{m}]\,\mathrm{MHz}$, 
$F_\mathrm{c}=40R^3[\mathrm{nm}]L^{-2}[\mu\mathrm{m}]\,\mathrm{pN}$, and 
$\alpha=1.5\times10^{27}R^3[\mathrm{nm}]L^{-5}[\mu\mathrm{m}]\,\mathrm{eV.m}^{-4}$.

\bibitem{footnote:spinless}
For simplicity we consider here spinless fermions. Within the approximations of this 
paper, the spin would add only a multiplicity factor in the tunneling rates. 

\bibitem{maria09_PRB}
E.\ Mariani and F.\ von Oppen, Phys.\ Rev.\ B {\bf 80}, 155411 (2009).

\bibitem{sapma03_PRB}
S.\ Sapmaz, Ya.\ M.\ Blanter, L.\ Gurevich, and H.\ S.\ J.\ van der Zant, 
Phys.\ Rev.\ B \textbf{67}, 235414 (2003).

\bibitem{izumi05_NJP}
W.\ Izumida and M.\ Grifoni, 
New J.\ Phys.\ \textbf{7}, 244 (2005).

\bibitem{sapma06_PRL}
S.\ Sapmaz, P.\ Jarillo-Herrero, Ya.\ M.\ Blanter, C.\ Dekker, and H.\ S.\ J.\ van der Zant, 
Phys.\ Rev.\ Lett.\ \textbf{96}, 026801 (2006).

\bibitem{flens06_NJP}
K.\ Flensberg, 
New J.\ Phys.\ \textbf{8}, 5 (2006).

\bibitem{footnote:coupling}
More precisely, if $gX^2n_\mathrm{d}/2$ is the intrinsic coupling term (see
Ref.~\onlinecite{weick10_PRB}), we 
can neglect this term with respect to the linear one if 
$g \ll F_\mathrm{e}^{2/3} \alpha^{1/3}$.

\bibitem{husse10_preprint}
R.\ Hussein, A.\ Metelmann, P.\ Zedler, and T.\ Brandes, 
Phys.\ Rev.\ B \textbf{82}, 165406 (2010).

\bibitem{weiss}
U.\ Weiss, 
\textit{Quantum Dissipative Systems} 
(World Scientific, Singapore, 1993)

\bibitem{mohan02_PRB}
P.\ Mohanty, D.\ A.\ Harrington, K.\ L.\ Ekinci, Y.\ T.\ Yang, M.\ J.\ Murphy,
and M.\ L.\ Roukes, 
Phys.\ Rev.\ B \textbf{66}, 085416 (2002).

\bibitem{seoan07_EPL}
C.\ Seoanez, F.\ Guinea, and A.\ H.\ Castro Neto, 
EPL \textbf{78}, 60002 (2007).

\bibitem{seoan08_PRB}
C.\ Seoanez, F.\ Guinea, and A.\ H.\ Castro Neto, 
Phys.\ Rev.\ B \textbf{77}, 125107 (2008).

\bibitem{seoan07_PRB}
C.\ Seoanez, F.\ Guinea, and A.\ H.\ Castro Neto, 
Phys.\ Rev.\ B \textbf{76}, 125427 (2007).

\bibitem{elste08_APA}
F.\ Elste, G.\ Weick, C.\ Timm, and F.\ von Oppen, 
Appl.\ Phys.\ A \textbf{93}, 345 (2008).

\bibitem{footnote:fluctuation-dissipation}
Notice that at
equilibrium ($V=0$), the fluctuating and dissipative parts of the Fokker-Planck
equation \eqref{eq:FP} fulfills the fluctuation-dissipation theorem, i.e.,
$[D(X)/2+\eta_\mathrm{e}k_\mathrm{B}T]/[\eta(X)+\eta_\mathrm{e}]=k_\mathrm{B}T$.

\bibitem{footnote:adiabatic}
Notice that in actual experiments on suspended carbon nanotube quantum dots, the ratio
$\omega_0/\Gamma$ is typically very small, of the order of $10^{-2}$ (see 
Refs.~\onlinecite{steel09_Science}, \onlinecite{lassa09_Science}, \onlinecite{sapma06_PRL}).

\bibitem{weick_unpublished}
G.\ Weick \textit{et al.}, unpublished.

\bibitem{chaikin}
P.\ M.\ Chaikin and T.\ C.\ Lubensky, 
\textit{Principles of Condensed Matter Physics} 
(Cambridge University Press, Cambridge, 1995).

\bibitem{footnote:Teff}
At equilibrium ($v=0$), the effective temperature \eqref{eq:T_eff} corresponds
to the temperature of the electron reservoirs $T$, cf.\
Ref.~\onlinecite{footnote:fluctuation-dissipation}.

\bibitem{footnote:hot}
Notice, however, that this does not imply a heating of the electron
reservoirs as these are considered to be macroscopic. Thus, there is no
back-action of the mechanical degree of freedom on the electronic leads which
remain at the equilibrium temperature $T$.

\end{thebibliography}
\end{document}